\documentclass{IEEEtran}
\usepackage{cite}
\usepackage{amsmath,amssymb,amsfonts}
\usepackage{algorithmic}
\usepackage{textcomp}
\usepackage[utf8]{inputenc}
\usepackage[T1]{fontenc} 
\usepackage{hyperref} 
\usepackage{mathtools}
\usepackage{graphicx}
\usepackage{tabularx}
\usepackage{subfigure}
\usepackage[table]{xcolor}
\graphicspath{{images/}}
\usepackage{dblfloatfix}
\usepackage{fixltx2e}
\usepackage{multirow}
\usepackage[justification=centering]{caption}
\usepackage{array}						
\usepackage{booktabs}					
\usepackage{float}
\usepackage{multirow}
\usepackage[table]{xcolor}

\usepackage{mathrsfs} 				
\usepackage{amsthm}			

\usepackage{soul}

\begin{document}

\title{The irruption of cryptocurrencies into Twitter cashtags: a classifying solution}

\author{\IEEEauthorblockN{Ana Fern\'andez Vilas}
\IEEEauthorblockA{AtlantTTic Research Center, University of Vigo, Spain.
Email: avilas@det.uvigo.es}\\
\and
\IEEEauthorblockN{Rebeca Díaz Redondo}
\IEEEauthorblockA{AtlantTTic Research Center, University of Vigo, Spain.
Email: rebeca@det.uvigo.es}\\
\and
\IEEEauthorblockN{Ant\'on Lorenzo Garc\'ia}
\IEEEauthorblockA{AtlantTTic Research Center, University of Vigo, Spain}}

\maketitle

\begin{abstract}

There is a consensus about the good sensing characteristics of Twitter to mine and uncover knowledge in financial markets, being considered a relevant feeder for taking decisions about buying or holding stock shares and even for detecting stock manipulation. Although Twitter hashtags allow to aggregate topic-related content,  a specific mechanism for financial information also exists: Cashtag (consisting of the company  ticker preceded by  \$) is a supporting mechanism  to track financial tweets referring to a company listed in a stock market.  However, according to our experiments and due to the  lack of conventions in cashtags usage, the irruption of cryptocurrencies has resulted in a significant degradation on the cashtag-based aggregation of posts. Unfortunately,  Twitter' users may use homonym tickers to refer to cryptocurrencies and to companies in stock markets, which means that filtering by cashtag may result on both posts referring to stock companies and  cryptocurrencies. This research proposes  automated classifiers to distinguish conflicting cashtags and, so, their container tweets by analyzing the distinctive features of tweets referring to stock companies and cryptocurrencies. As experiment, this paper analyses the interference between cryptocurrencies and company tickers in the London Stock Exchange (LSE), specifically, companies in the main and alternative market indices FTSE-100 and AIM-100. Heuristic-based as well as supervised classifiers are proposed and their advantages and drawbacks, including their ability to self-adapt to Twitter usage changes, are discussed. The experiment confirms a significant distortion in collected data when colliding or homonym cashtags exist, i.e., the same  \$ acronym to refer to company tickers and cryptocurrencies. According to our results, the distinctive features of posts including cryptocurrencies or company tickers support accurate classification of colliding tweets (homonym cashtags) and Independent Models, as the most detached classifiers from training data, have the potential to be trans-applicability (in different stock markets) while retaining performance.
\end{abstract}


\section{Introduction }
\label{sec:introduction}
The increasingly irruption of information \& telecommunications technologies in the stock markets have been accompanied by business growth. The flood of information can support decisions taken by brokers as well as individual investors which harvest information about the situation of a company,  clients' opinions, socio-economical changes, political decisions, rumors, etc. Moreover, the ubiquity of online social media has caught also companies, brokers and other key roles in the financial market which have begun to share more and more financial information and expert opinions on stock exchanges. Although all this information might turn social media into one of the main sources -if not the main one- for decision-making due to its real-time nature, success in stock trade  depends not only in quick access to information but also on the quality of this information.\\

\noindent Currently, Twitter is one of the most used platforms to share financial information from companies, brokers, news agencies or individual investors. Above other financial information sources like message boards or discussion forums, stock microblogging exhibit three distinctive characteristics  \cite{Sprenger2014}: (1) Twitter's public timeline may capture the natural market  conversation more accurately and reflect up to date developments;  (2) Twitter supports a more ticker-like live conversation, which  allows twitter-microbloggers to be exposed to the most recent information of  all stocks and does not require users to actively enter the forum for a  particular stock; and (3) twitter-microbloggers should have a stronger incentive to publish valuable information in order to maintain reputation (increase mentions, the rate of retweets and their followers), while financial  bloggers can be indifferent to their reputation in the forum. The combinations of this stream of information with suitable processing and analysis techniques would support the action for many financial stakeholders and even law enforcement agencies.\\

\noindent The main Twitter sharing mechanism to track financial information is the cashtag: a clickable term consisting in a company ticker preceded by \$ symbol. Remember that a company ticker  is a short sequence of letters and sometimes numbers, that identifies uniquely a company in a specific stock market. For example, in the case of Vodafone, its ticker in LSE (London Stock Exchange) is VOD so that  its cashtags would be \$VOD. This cashtag is included in the tweet's text similarly to what happens with hashtags. It is supposed that the usage of \$VOD  marks the tweet as a post containing financial information about Vodafone. Also similarly to hashtags, Twitter supports tracking tweets that contain a specific cashtag. All of this turns cashtags into one of the most useful mechanisms to easily harvest financial information on Twitter. However, the  irruption of  cryptocurrencies  has degraded the accuracy and so the quality  of the information obtained through cashtags because some cryptocurrencies acronyms are equal to company tickers in stock markets (acronym conflict), largely due to the huge quantity of cryptocurrencies  and the lack of a cryptocurrencies' regulated market.  As a result, when a conflicting or colliding cashtag is used for tracking or searching, results referring to both stock companies and cryptocurrencies might be retrieved. Moreover, (i)  the total amount of cryptocurrency-related tweets extensively surpasses  company-related tweets (as a consequence of the increasing  popularity of cryptocurrencies) and (ii)  cryptocurrency tweets are considered low quality since most of them are spam or auto-generated messages. All of this produces a very significant degradation in the tracking capacity of cashtags and underlines the need of disambiguation mechanisms to distinguish both groups of  tweets.\\

\noindent This paper aims to highlight the conflicting issue in cashtags as well as proposing classifiers to distinguish cryptocurrency and company cashtags with enough accuracy. To show the challenge and introduce the classifiers, the research work was deployed over real data retrieved from Twitter and related to the London Stock Exchange (LSE). The experiment in this paper focus on cashtag conflicts in LSE companies, specifically, the 100 companies listed in the main market index  FTSE-100 and the 100 companies listed in the alternative market index AIM-100,  both indices during the period July 1, 2017 and February 15, 2018. Hereinafter, we refer as LSE-100 to both indices. \\

\noindent  This paper is structured as follows. After introducing the related work and motivation in Section \ref{sec:related}, the motivation, the experimental dataset and the exploratory analysis is described in Section  \ref{sec:motivation}, where the impact of cryptocurrency tweets on the LSE-100  tweets is quantified. The datasets  and  a  bird-eye description of the  deployed methodology  are introduced in Sections  \ref{Datasets} and \ref{sec:methodology}. Then, exploratory analysis, where distinctive characteristics are uncovered for  LSE-100 tweets  (tweets that only refer to a company listed in LSE-100) and cryptocurrency-tweets (tweets that only refer to a cryptocurrency) in a process of feature extraction is detailed in Section \ref{sec:tweet_features}.  From the mentioned methodology, the proposed classifying systems, which solve the problem of colliding cashtags, are progressively introduced:   Word-based Heuristic Filters (Section \ref{sec:filters}); SVM (Support Vector Machine) Classifiers (Section  \ref{SVM Classifiers}); Combined Classifiers (Section \ref{Combined Classifiers}); LSTM (Long short-term memory) network classifiers (Section \ref{LSTM Classifiers}); and Logistic-regression-based Classifiers (Section \ref{Logistic-regression-based Classifiers}). The different advantages and drawbacks of the proposed classifiers, as well as their ability to self-adapt to Twitter usage changes, are discussed in Section \ref{limitations} and, given the generalizable features of the independent versions of the classifiers, Section \ref{modelEvaluation} applies a statistical test to evaluate if the different performances are due to a difference in the models. Finally, the main conclusions and future work are summarized in Section \ref{conclusions}.\\

\section{Related Work }
\label{sec:related}

The modernization and digitalization of the financial market has been accompanied by the remarkable increasing of online information available for both brokers and individual investors, especially in social media, which bring up a huge source of knowledge which may be applied to  analysis and even predict financial movements and so to assist in taking decisions in financial markets. Several works have been accomplished regarding the predictive power of the financial information  in social media.   \cite{10.1371/journal.pone.0040014} show that trading volumes of stocks listed in NASDAQ-100 are correlated with their query volumes, the number requests submitted on the Internet, and \cite{WANG2012136} proposed sentiment analysis as one of the most relevant features to improve the accuracy of financial time-series forecasting. More recently, \cite{CAVALCANTE2016194} and \cite{Li2017}  raised similar ideas such as the importance of mining textual context and sentiment analysis of professional opinions in social media and financial news as useful supplementary sources for forecasts; and \cite{8481411} applies regression and time series models with social media data ({\it Twitter})  and stock market values to predict monthly total vehicles sales.\\

\noindent In addition, several intelligent trading systems have been proposed, such as \cite{GUNDUZ20159001} that deployed a forecasting method which combines the analysis of news  from Turkish finance websites, the extraction of feature vectors and the stock prices to predict future market movements; or \cite{KHADJEHNASSIRTOUSSI2015306} which applied text mining techniques to financial news-headlines and predict movements in the FOREX market. Deep learning has been also applied to model both short-term and long-term influences of events on stock price movements in \cite{ding2015deep}. Finally, in \cite{10.1371/journal.pone.0146576} the combination of public news with the browsing activity of the users of Yahoo! Finance to forecast intra-day and daily price changes of a set of 100 highly capitalized US stocks was explored. To sum up, most of the research in this field highlight the predictive power of social media, especially combined with others information sources.  In \cite{8464661}, the idea that stock markets are impacted by various factors (trading volume, news events and the investors' emotions) is supported via multi-source multiple instance learning applied to a 3-part dataset: historical quantitative data;
Web news articles; and investor social network posts (all located in China).

\noindent If the impact of information from online data sources into the financial market is widely acknowledged by researchers and professionals, there is also a huge consensus about Twitter specifically.  Besides the well-known hashtag, that allows people to follow topics they are interested in, Twitter unveiled a new clicking and tracking feature for tickers (companies' stock symbols) known as cashtags which are, as explained before. Cashtags allow  tracking financial information about a specific company or market. Related to this mechanism, \cite{Hentschel2014} reported an exploratory analysis of public tweets in English which contain at least one cashtag from NASDAQ ({National Association of Securities Dealers Automated Quotation}) or NYSE (New York Stock Exchange). The research concludes that the use of cashtag is higher in the technologic sector, which seems to be related to the technological profile of most of the Twitter users. It also highlights the existence of relevant information behind the co-occurrence of cashtags and the co-occurrence of cashtags and hashtags together.\\

\noindent It deserves to be mentioned that according to \cite{Dredze2016} there are mainly five types of users that post financial information in Twitter (journalist, companies and their representatives,  investors, government agencies and citizen journalists) and that, contrary to the classic information sources -consisting mainly of breaking news-, Twitter financial information also includes rumors and speculations.  In \cite{Ceccarelli2016}, authors  agree on the usefulness of Twitter as a source of financial information and in its complementary value to traditional information sources, but they suggest that, regarding Twitter popularity, is is not necessarily the same for financial contributors as for contributor in  other areas, so that, the importance of novelty and popularity is higher in financial tweets. Moreover, in \cite{elliott2017disclosure}, it is shown that the impact of negative financial news over investors depends also on its origin.   When the negative news comes from the Twitter account of the Investor Relations Office, the investors' willingness to invest highly decrease but when the negative news come from the CEO's Twitter account, they have no effect on it. \\

\noindent \cite{Rao2014} studied the relationship between Twitter sentiment and  financial market measures like volatility, trading volume, etc. with promising results in Dow Jones Industrial Average (DJIA) and NASDAQ-100  indices for high-frequency trades. This type of algorithmic trading --high speed, high turn over rate -- demands real-time financial data and electronic trading tools, and especially social media could be integrated as a fast mechanism to capture the public behavior and opinion to take  decisions. More recently, in  \cite{Vilas2008Twitter}  it is analyzed the variability on  posts' volume, content, sentiment and geographical provenance after a far-impacting financial event  and concludes that although Twitter is not a specific-purpose financial forum, it is highly permeable to financial events. Thus, post's sentiment changes with the considered financial event so that Twitter activity  can be a good predictor or sign of the stock market state. However, not everything related to Twitter is good news. According to \cite{Dredze2016}, the use of Twitter adds new challenges like the huge volume of available data, the high level of  repetition of the same information or the quality of the tweets.\\

\noindent With all the above, the relationship between Twitter behavior and stock share price is, by far, the most studied scenario, especially in  relevant moments as quarterly announcements. In \cite{Ranco2015} a 15-month period of Twitter data about 30 stock companies  of the DJIA index was investigated and, according to the results, it can be stated that not only is there a strong correlation between Twitter behavior and stock share price in well known relevant moments, but also there are correlation peaks which do not correspond to any expected news about the stock market. Moreover, in \cite{Liu2015}, Twitter is used to identify and predict stock co-movement from firm-specific social media metrics. Aside from causality between Twitter and stock prices, in \cite{Shutes2016}  US market tweets  are studied as signs of new information in the stock market and the experiment shows that nearly a third of the tweets are linked to abnormal price movements. However, the lack of information during regular periods makes difficult that Twitter completely replace traditional information sources for the financial market. \\

\noindent Leaving apart  the relation between Twitter and financial markets, other researches have studied the predictive value of the information extracted from Twitter to take trading  decisions. In \cite{ruiz2012correlating} the correlation between Twitter activity and financial time-series showed that stock share prices are weakly correlated with the analyzed Twitter features if they are used alone.  In addition, \cite{7838051},  analyzed over 1700 listed companies for more than two years. Apart from the importance of obtaining a huge financial tweet dataset, the authors found out that expert users impact the financial market more than others and that  technology and consumers show a better correlation than other sectors.\\

\noindent Even though most of the research work focused on Twitter data volume, as the ones previously introduced, some studies also apply sentiment analysis to distinguish the polarity of Twitter content and its impact on the financial market. \cite{BOLLEN20111} showed that public mood analyzed through Twitter feeds is  correlated with the DJIA. Also, \cite{zhang2013sentiment}, found out a high negative correlation between mood states like hope, fear and worry in tweets and the DJIA. Furthermore, in 
\cite{al2014big}, \cite{Liew2016}, \cite{Rajesh2016}, \cite{Cortez2016} and  \cite{nguyen2015sentiment}  sentiment analysis was considered useful to make trading decisions or predict  stock market variables. More recently, \cite{pagolu2016sentiment} applied sentiment analysis and unsupervised machine learning to analyze the correlation between stock market movements of a company and sentiments in tweets, finding out a strong correlation between the rises and falls in stock prices of a company and public opinions or emotions about that company expressed on Twitter. Also, \cite{dickinson2015sentiment} investigated the Pearson correlation of public sentiment with stock increases and decreases. Also, \cite{OLIVEIRA201662, OLIVEIRA2017125} proposed stock market lexicons to deal with the short length of tweets, one of the main issues of natural language techniques when they are applied to tweets. Then, they studied the correlation between investors' sentiment indicators and two traditional survey-based indicators --II (Investors Intelligence) and AAII (American Association of Individual Investors) with moderate correlation results.\\

\noindent To sum up, there is a consensus of the good sensing and novelty characteristics of Twitter as a source of information for the financial market, especially if it is combined with other information sources. As most of the current research is focused on the predictive power of Twitter  and on their capability to support decision making, now, it is especially important to recover information with enough quality to support these foreseen expert systems for the financial market. At this respect, this paper aims to support quality in financial data retrieving. This research work highlights the negative effect of the  popularity of cryptocurrencies in the sensing capability of Twitter, and specifically on the efficiency of cashtags as a tracking mechanism for financial information due to, as mentioned before,  the usage of homonyms cashtags to refer to both company tickers and cryptocurrencies.

\section{Motivation }
\label{sec:motivation}

Since 2012, Twitter incorporates cashtag as a mechanism to find and track tweets that address  companies by their tickets in a specific stock market. However its usefulness has been deteriorated due to the interference of cryptocurrencies. Although cryptocurrencies  have existed for a long time, they became remarkably popular at late 2017 as it is shown in Figures 1(a), 1(b) and 1(c) which represent the volume of Google searches pro the term "cryptocurrency" and two specific ones {\it Nxt} and {\it Stellar Lumens}. In the same period, the volume of Google searches about specific stock companies that have remain constant, Figure 1(d).  \\

\begin{figure}[t!] 
\begin{tabular}{c}
\subfigure[\textit{Cryptocurrency} (general term) Searches on Google trend evolution]
    {
        \includegraphics[width=\columnwidth]{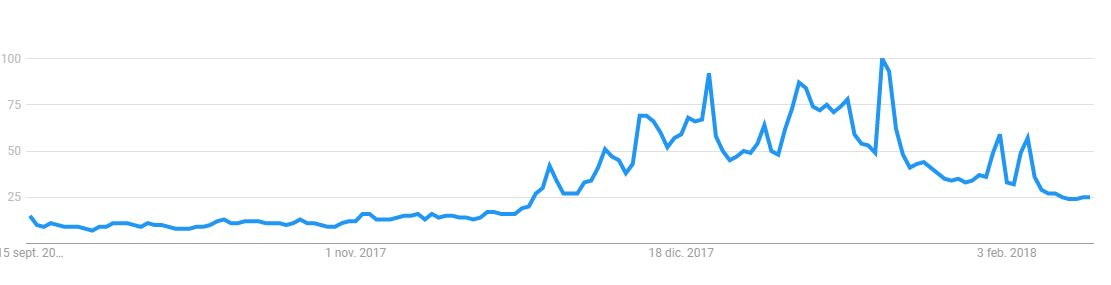}

   }
\\
  \subfigure[\textit{NXT} (ticker of Nxt platform, cryptocurrency) Searches on Google trend evolution]
    {
        \includegraphics[width=\columnwidth]{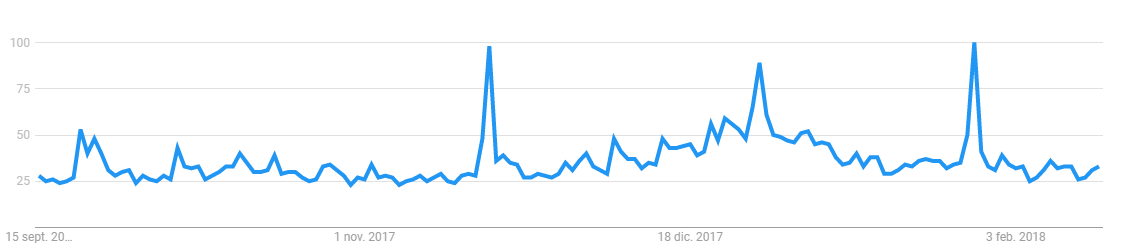}
   }
\\
   \subfigure[\textit{XLM} (ticker of Stellar Lumens, cryptocurrency) Searches on Google trend evolution]
    {
        \includegraphics[width=\columnwidth]{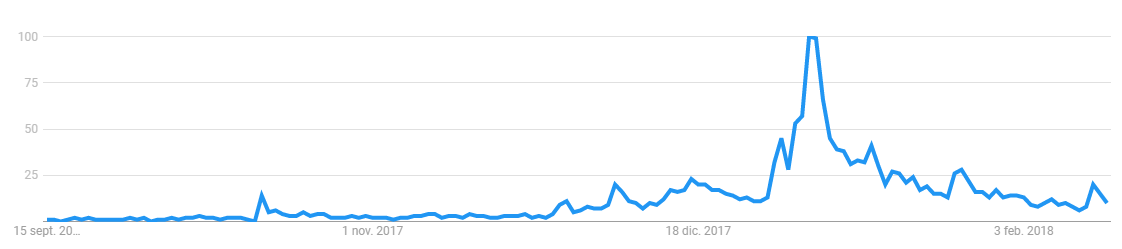}
   }
   \\
   \subfigure[\textit{VOD} (ticker of Vodafone company in  LSE) Searches on Google trend evolution]
    {
        \includegraphics[width=\columnwidth]{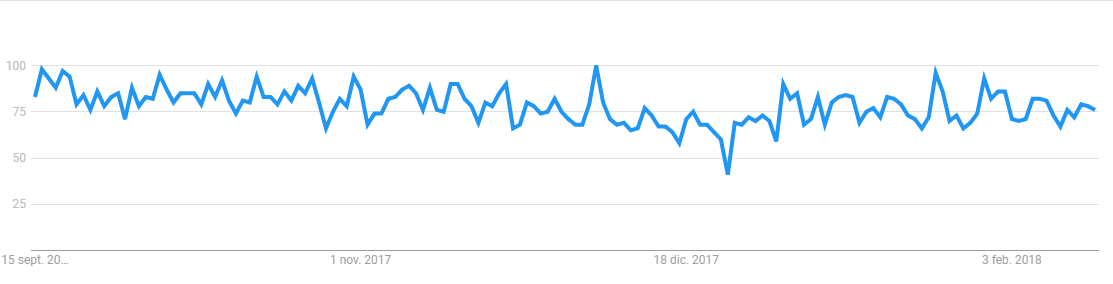}
   }\\
\end{tabular}
  \caption{Searches on Google trend evolution late 2017}
\end{figure}

\noindent This change in behavior is also visible on Twitter, where the number of daily results about cryptocurrencies has increased by more than 40 times, according to our analysis. Although these tweets should not interfere with cashtag mechanisms, many of them use the dollar symbol \$ followed by the acronym of the cryptocurrency to indicate that the tweet refers to it. The conflict arises when the ticker of some company and the acronym of a certain cryptocurrency match, what is a natural consequence of the huge number of cryptocurrencies that have emerged in a short time. As a result, it may happen that, at recovering tweets with a specific cashtag, most of them do not refer to the company they should identify, but they address the coincident cryptocurrency instead. As an example, this is the case for \$XLM ({\it XLMedia} company  vs {\it Stellar Lumens} cryptocurrency) and \$NXT ({\it Next plc} company vs {\it Nxt platform} cryptocurrency). We refer to these colliding cashtags as \textit{homonym cashtags} and the tweets that contain at least one of them \textit{homonym tweets}.\\

\noindent As mentioned, this paper studied the negative effect of homonym cashtags by using LSE-100 as study stock exchange. So, a {\it homonym cashtag} is any cashtag that can refer to both an LSE (LSE-100, restricted to companies in FTSE-100 and AIM-100) company and a cryptocurrency, because both have the same acronym and a {\it homonym tweet} is  any tweet that has at least one homonym cashtag. The list of {\it homonym tickers} in LSE-100 can be seen in Table \ref{table:homonym:tickers}. These tickers were identified manually, looking for coincident cryptocurrencies for each constituent company. On the other hand, we will call {\it non-homonym cashtag} to any cashtag that only refers to a regulated stock market company or to a cryptocurrency, because there are not two of them with the same ticker, and {\it non-homonym tweet} to any tweet that has at least one cashtag from an LSE-100 company, as long as none of its tickers is included in the list of homonym cashtags. We consider a \textit{company tweet} any tweet that contains at least one cashtag that refers to a company in a stock market and \textit{cryptocurrency tweet} to any tweet that contains at least one cashtag that refers to a cryptocurrency.\\

\begin{figure}[ht] 
  \centering
  \scalebox{1}[1]{ \includegraphics[width=\columnwidth]{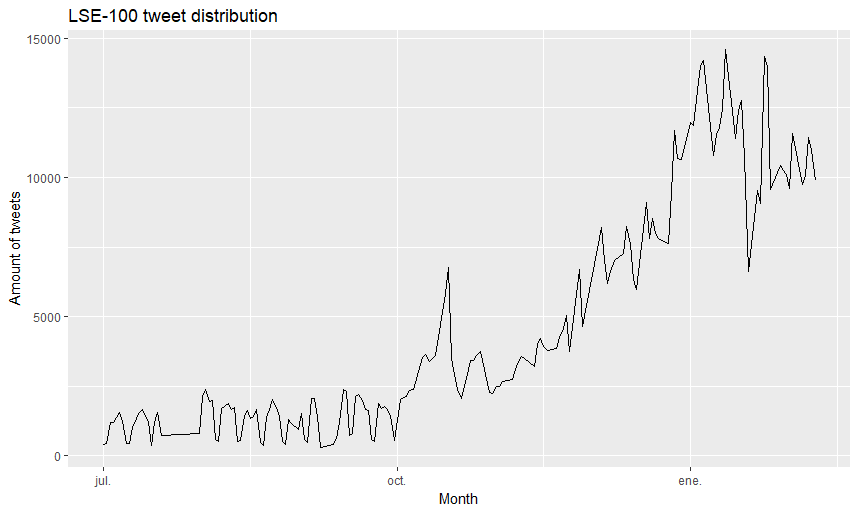}}
  \caption{FTSE-100 and AIM-100 tweets: time distribution (July 2017-February 2018)}
	 \label{fig:ftsetimedistribution}
\end{figure}

\begin{table}[ht]
	{\footnotesize
\centering
\begin{tabular} {lll}
  \hline
\begin{minipage} {1.5 cm}   \textbf{Homonym cashtags} \end{minipage} & \begin{minipage} {3.5 cm}  \textbf{LSE-100 company(market)}  \end{minipage} & \begin{minipage} {2.5 cm}  \textbf{Cryptocurrency} \end{minipage} \\ 
  \hline
\begin{minipage} {1.5 cm}  \$NXT  \end{minipage} & \begin{minipage} {3.5 cm} Next plc (FTSE-100)  \end{minipage} & \begin{minipage} {2 cm}   Nxt (coin and platform)  \end{minipage} \\ 
\hline
\begin{minipage} {1 cm} \$SKY  \end{minipage} & \begin{minipage} {3.5 cm}  SKY plc (FTSE-100)  \end{minipage} & \begin{minipage} {2 cm}   Skycoin \end{minipage} \\
\hline
\begin{minipage} {1.5 cm}  \$XLM \end{minipage} & \begin{minipage} {3.5 cm}  XLMEDIA (AIM-100) \end{minipage} & \begin{minipage} {2 cm}    Stellar  \end{minipage} \\
\hline
\begin{minipage} {1.5 cm} \$BRK \end{minipage} & \begin{minipage} {3.5 cm}  BROOKS (AIM-100)  \end{minipage} & \begin{minipage} {2 cm}   Breakout coin  \end{minipage} \\
\hline
\begin{minipage} {1.5 cm}  \$GBG  \end{minipage} & \begin{minipage} {3.5 cm}  GB group  (AIM-100)  \end{minipage} & \begin{minipage} {2 cm}    Golos Gold  \end{minipage} \\
\hline
\begin{minipage} {1.5 cm}  \$APH  \end{minipage} & \begin{minipage} {3.5 cm}  Alliance pharma (AIM-100)  \end{minipage} & \begin{minipage} {2 cm}   Aphroditecoin  \end{minipage} \\
\hline
\begin{minipage} {1.5 cm}  \$AMS  \end{minipage} & \begin{minipage} {3.5 cm}  Advanced medical solutions (AIM-100)  \end{minipage} & \begin{minipage} {2 cm}   Amsterdamcoin  \end{minipage} \\
\hline
\begin{minipage} {1.5 cm}  \$CRW \end{minipage} & \begin{minipage} {3.5 cm}  Craneware (AIM-100)  \end{minipage} & \begin{minipage} {2 cm}  Crown  \end{minipage} \\
\end{tabular}}
\caption{Homonym tickers in LSE (LSE-100, restricted to FTSE-100 and AIM-100) \label{table:homonym:tickers}}
\label{tab:homonymtickers}
\end{table}

\begin{figure}[ht] 
  \centering \includegraphics[width=\columnwidth]{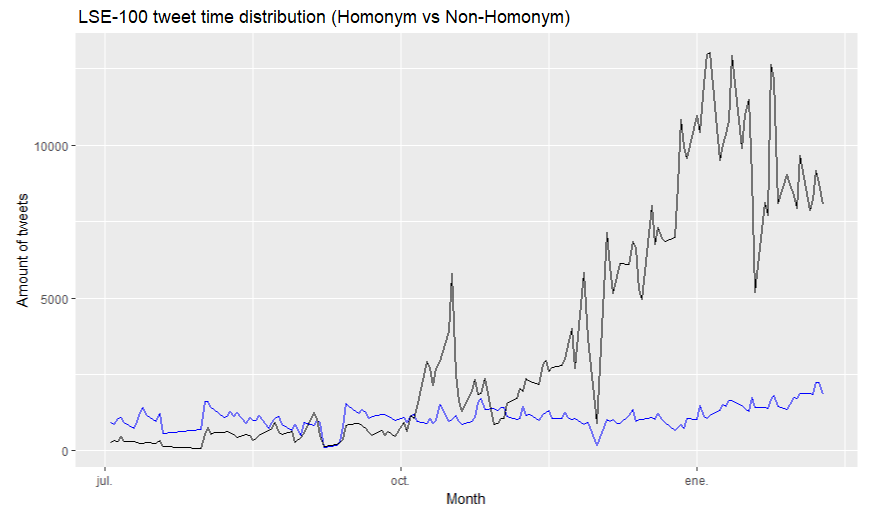}
  \caption{Daily FTSE-100  and AIM-100 tweet time distribution, \textit{Homonym}(black) vs \textit{No Homonym}(blue) (July 2017-February 2018) \label{Homonym-non-momony-tweet-distribution} }
\end{figure}

\noindent The number of tweets that contain a FTSE-100 or an AIM-100 cashtag sharply  increased at late 2017 (Figure \ref{fig:ftsetimedistribution}). However, most of the tweets do not refer to LSE-100 companies. Remember that  FTSE-100 index lists  the one hundred most valuable companies  such as Vodafone, {\it Cocacola} or {\it RioTinto}, while the AIM-100 lists the one hundred most valuable companies in the secondary market, these companies (eg. {\it Alliance Pharma}, {\it Hutchison China} or {\it Stafflineare}) less known than the main market companies. This interference is even more impacting if we take into account the disparate number of results obtained. While looking at FTSE-100 non-homonym tickers  we have up to 1,000 results daily, the number of daily results that refer to the XLM and NXT (Cryptocurrency-colliding tickets) are more than 10,000 (Figure \ref{Homonym-non-momony-tweet-distribution}). \\ Apart from that, as it is shown in Figure \ref{Homonym-non-momony-tweet-distribution}, the interference of homonym tweets skyrocketed more than 30 times since October 2017. From this period  the recovered homonym tweets made up practically all the results obtained. In fact, the number of recovered homonym tickers  in December are 5.6 times the amount collected for the non-homonym tweets  for the FTSE-100 market and up to 40 times for the AIM-100.\\

\begin{figure}[ht] 
  \centering
  \scalebox{1}[0.7]{ \includegraphics[width=\columnwidth]{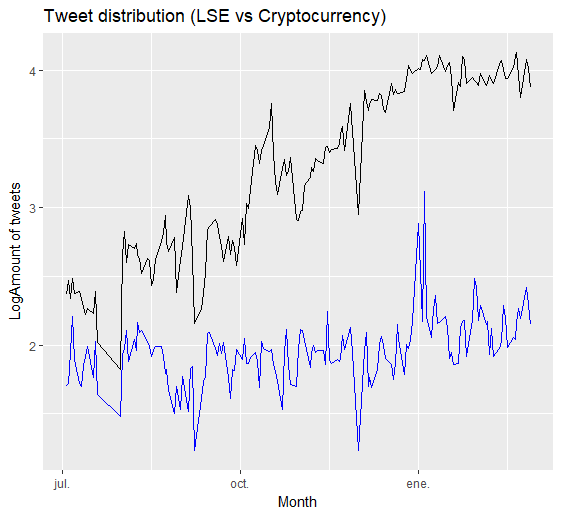}}
  \caption{\label{Homonym tweet distribution} Homonym tweets time distribution, \textit{LSE-100}(blue) vs \textit{Cryptocurrency}(black) (July 2017-February 2018)}
\end{figure}

\noindent To explore the interference of cryptocurrencies in regulated market, all the homonyms tweets were classified  manually as cryptocurrency tweet or LSE-100 tweet, taking into account the content, the user characteristics and history, etc. The results of the annotation are shown in Figure \ref{Homonym tweet distribution}, where we can see that the increase in homonym tweets is mostly localized in cryptocurrency tweets, while the number of LSE-100 tweets remains constant, with a ratio 100:1 at the beginning of 2018. 

This large number of cryptocurrency tweets underline the need of support methods to properly retrieve information regarding stock exchanges. Although the situation varies slightly from one homonym cashtag to another, there is a clear sign of the difficulty to track the stock exchange via Twitter just by cashtags. In addition, almost all the tweets about cryptocurrencies are spam or auto-generated by applications. For this reason, the informative purpose of the cashtag is almost lost, so some disambiguation mechanisms need to be developed.

\section{Datasets }
\label{Datasets}

To carry out this paper, three different datasets have been used according to wether they include or not a set of cashtags we selected to have a representative set of non-homonym cryptocurrency cashtags, non-homonym LSE-100 cashtags (FTSE-100, AIM-100) and homonym cashtags:
\begin{itemize}
	\item {\it Non-homonym  Cryptocurrencies tweets}: a set of tweets that contains at least one of the cashtags of  non-homonym cryptocurrencies, that is, no coincident with the tickers in  FTSE-100 and AIM-100. 
\item {\it Non-homonym LSE-100 tweets}: a set of  tweets that have at least one cashtag from an LSE-100 company, as long as it does not contain and  homonym cashtag, respectively split into two subsets (FTSE-100 and AIM-100). Keep in mind that a tweet can be in both subsets if it has a cashtag from  FTSE-100 and a cashtag from  AIM-100. Likewise, if the tweet also has a  non-homonym cryptocurrency cashtag, the tweet could belong to {\it Cryptocurrencies non-homonym tweets}. 
\item {\it Homonym tweets:} a set of tweets that contains at least one homonym cashtag which refers both to a company and to a cryptocurrency. Also, it can be split into two subsets, FTSE-100 and AIM-100 according to the list in  which the colliding cashtag is included. 
\end{itemize}

\begin{table}[ht]
\centering
\begin{tabular} {|l|}
 \hline
{\bf Cryptocurrencies} \\
\hline
\hline
\begin{minipage}[t]{\columnwidth}
{\footnotesize
\$SNT, \$ADA, \$MTH, \$ADX, \$LSK, \$DSR, \$ARK, \$CLOAK, \$TKN, \$DLC, \$DCR, \$KMD, \$IQT, \$ZCL, \$DCY, \$ALIS, \$RBY, \$SYS, \$EXP, \$BCY, \$VEN, \$BCN, \$BLITZ, \$UGT, \$GVT, \$MONA, \$QASH, \$DASH, \$AUR, \$UNO, \$BURST, \$REQ, \$PART, \$TRIG, \$GCR, \$LMC, \$XEM, \$BNB, \$SNGLS, \$BITSILVER, \$PDC, \$ELIX, \$XVG, \$DOPE, \$LEND, \$SNRG, \$NLG, \$ARDR, \$QSP, \$SALT, \$SYNX, \$GRC, \$XDN, \$PIVX, \$DCT, \$WAVES, \$PTOY, \$SIB, \$LTC, \$CPC, \$NAS, \$XMR, \$LOCI, \$ION, \$VSX, \$NXS, \$XMY, \$GBYTE, \$XMG, \$BAT, \$IOP, \$HMQ, \$NTCC, \$PKB, \$BAY, \$PBL, \$BYC, \$MINT, \$HSR, \$MUSIC, \$XSPEC, \$IGNIS, \$ETP, \$BWK, \$FCT, \$DRGN, \$MUE, \$XPM, \$STEEM, \$FTC, \$SPHR, \$DGB, \$DGD, \$SUB, \$VOX, \$MAID, \$RPX, \$AEON, \$XAUR, \$MIOTA, \$CRC, \$BET, \$ENG, \$XVJ, \$POWR, \$STORJ, \$GUP, \$UBQ, \$SBD, \$INFX, \$LGD, \$DYN, \$INFR, \$ONION, \$MANA, \$SLR, \$FUN, \$CURE, \$BITB, \$EMC2, \$XZC, \$IOTA, \$COVAL, \$AGRS, \$PASC, \$DOGE, \$XRB, \$SWT, \$FLDC, \$ZEC, \$NBT, \$XRP, \$ETH, \$RADS, \$ETC, \$PANGEA, \$CLAM, \$PHR, \$APX, \$BTC, \$NEM, \$NEO, \$MYST, \$START, \$ENJ, \$WTC, \$PPT, \$STR, \$ARDOR, \$ITZ, \$BCPT, \$ITC, \$TAAS, \$STRAT, \$SEQ, \$EDG}
\end{minipage}\\
\hline
\hline
{\bf FTSE-100 }\\
\hline
\hline
 \begin{minipage}[t]{\columnwidth}
{\footnotesize
\$CPI, \$DC., \$HIK, \$INTU, \$SN., \$CPG, \$CCL, \$BARC, \$CCH, \$GSK, \$BDEV, \$DCC, \$BLND, \$RIO, \$WTB, \$SMIN, \$IAG, \$MRW, \$SVT, \$III, \$ITRK, \$AHT, \$JMAT, \$IHG, \$LGEN, \$HL., \$AV., \$BATS, \$STAN, \$CRH, \$LSE, \$RTO, \$SGRO, \$SBRY, \$CRDA, \$SHP, \$DLG, \$BLT, \$PSON, \$GKN, \$GLEN, \$PSN, \$NG., \$SSE, \$INF, \$SMT, \$BNZL, \$UU., \$MERL, \$REL, \$PRU, \$LAND, \$FERG, \$DGE, \$MDC, \$WPP, \$MCRO, \$EXPN, \$WPG, \$RRS, \$VOD, \$RMG, \$RR., \$IMB, \$RDSB, \$RDSA, \$HMSO, \$FRES, \$ADM, \$TSCO, \$PFG, \$HSBA, \$SKG, \$OML, \$TUI, \$ITV, \$MKS, \$ULVR, \$AZN, \$AAL, \$BT.A, \$BAB, \$PPB, \$BRBY, \$MNDI, \$RB., \$SL., \$LLOY, \$SGE, \$ABF, \$RBS, \$STJ, \$ANTO, \$CNA, \$SDR, \$GFS, \$TW., \$RSA, \$BP., \$CTEC, \$EZJ, \$KGF, \$BA. }
\end{minipage}\\
\hline
\hline
{\bf AIM-100}\\
\hline
\hline
\begin{minipage}[t]{\columnwidth}{\footnotesize \$OPG, \$SQS, \$PAF, \$BOO, \$GHH, \$TAP, \$MANX, \$SAA, \$VNL, \$KWS, \$IOM, \$PLUS, \$HZD, \$ARBB, \$BNN, \$IPEL, \$CVSG, \$SFE, \$OCI, \$CRS, \$DTG, \$STAF, \$FDP, \$ABC, \$XSG, \$SCH, \$BUR, \$BMK, \$APGN, \$TCM, \$HUR, \$NFC, \$IGR, \$SOLG, \$YNGN, \$FOG, \$QXT, \$REDD, \$YNGA, \$HCM, \$MPE, \$BREE, \$FEVR, \$RNWH, \$RWS, \$WINE, \$POLR, \$DOTD, \$SMS, \$TEF, \$GAMA, \$CLIN, \$MUL, \$CTH, \$TMO, \$JSG, \$CAM, \$ASY, \$QTX, \$ASC, \$NUM, \$CAKE, \$EMIS, \$LTG, \$SMTG, \$MAB1, \$VTU, \$JHD, \$CVR, \$PRSM, \$IQE, \$AMER, \$EAH, \$WJG, \$ACSO, \$SOU,\$CAML, \$JOUL, \$PANR, \$RTHM, \$FPM, \$MTW, \$VCP, \$HGM, \$DDDD, \$SLE, \$YOU, \$PURP, \$IDOX, \$OGN, \$HOTC, \$NICL, \$RST, \$MIDW, \$TFW, \$SCPA, \$SPH } 
\end{minipage} \\
\hline
\hline
\end{tabular}
\caption{Non-homonym Cashtags for the datasets \label{Cashtags-datasets}}
\end{table}

In Table  \ref{Cashtags-datasets} the three cashtag sets are shown. The set  of {\it Non-homonym  Cryptocurrencies }  was obtained from  web sites devoted to tracking cryptocurrencies. The set of {\it Homonym tickers} is shown in Table \ref{table:homonym:tickers} and it is the main scenario  analyzed in this paper: the incidence of new tweets related to cryptocurrencies on the tweets referring to a company in a stock exchange (LSE-100 in our case).  As mentioned, tweets  were manually annotated (cryptocurrency or company)  by expert considering their content, the user's characteristics and any additional information added to the tweet by a hyperlink were carefully analyzed. A schematic description of the  datasets is shown in Table \ref{datasets-overview}.\\

\begin{table*}[ht]
\footnotesize
\centering
{\footnotesize
\scalebox{1}[1]{
\begin{tabular} {llll}

\hline
\begin{minipage} {3 cm} \textbf{Name}  \end{minipage} & \begin{minipage} {4 cm}  \textbf{Data interval}  \end{minipage} & \begin{minipage} {5.5 cm}   \textbf{Description}  \end{minipage} & \begin{minipage} {1.5 cm} \vspace{1 mm} \textbf{Number of results} \vspace{1 mm} \end{minipage} \\ 
	
\hline
\hline
\begin{minipage} {3 cm}  Cryptocurrencies non-homonym tweets \textit{CNHDS}\end{minipage} & \begin{minipage} {4 cm} From 15 Jan 2018, to 15 Feb 2018 \end{minipage} & \begin{minipage} {4.5 cm}   \vspace{2.5 mm} Tweets that contain a cashtag of one of the main cryptocurrencies \vspace{2.5 mm} \end{minipage} & \begin{minipage} {1.5 cm}  1,023,232 \end{minipage} \\ 
	
\hline
\begin{minipage} {3 cm} FTSE-100 homonym tweets \textit{FTHDS} \end{minipage} & \begin{minipage} {4 cm}  From 1 Jul 2017, to 15 Feb 2018  \end{minipage} & \begin{minipage} {5.5 cm}   \vspace{2 mm} Tweets that contain a cashtag of companies of the FTSE-100 coincident with a cryptocurrency \vspace{2 mm}  \end{minipage} & \begin{minipage} {1.5 cm} 292,864 \end{minipage} \\ 
	
\hline
\begin{minipage} {3 cm}  FTSE-100 non-homonym tweets \textit{FTNHDS} \end{minipage} & \begin{minipage} {4 cm}  From 1 Jul 2017, to 15 Feb 2018  \end{minipage} & \begin{minipage} {5.5 cm}  \vspace{2 mm} Tweets that contain a cashtag of companies of the FTSE-100 that do not coincide with a cryptocurrency \vspace{2 mm}  \end{minipage} & \begin{minipage} {1.5 cm} 144,787  \end{minipage} \\ 
\hline
\begin{minipage} {3 cm} AIM-100 homonym tweets \textit{AMHDS} \end{minipage} & \begin{minipage} {4 cm} From 1 Jul 2017, to 15 Feb 2018  \end{minipage} & \begin{minipage} {5.5 cm}   \vspace{2 mm} Tweets that contain a cashtag of companies of the AIM-100 coincident with a cryptocurrency \vspace{2 mm}  \end{minipage} & \begin{minipage} {1.5 cm} 405,625  \end{minipage} \\ 
\hline
\begin{minipage} {3 cm}  AIM-100 non-homonym tweets \textit{AMNHDS} \end{minipage} & \begin{minipage} {4 cm}  From 1 Jul 2017, to 15 Feb 2018  \end{minipage} & \begin{minipage} {5.5 cm}  \vspace{2 mm} Tweets that contain a cashtag of companies of the AIM-100 that do not coincide with a cryptocurrency \vspace{2 mm}  \end{minipage} & \begin{minipage} {1.5 cm} 69,138  \end{minipage} \\ 
\hline
\end{tabular}}
}
\caption{Overview of the datasets\label{datasets-overview}}
\end{table*}

\subsection{Tweet structure}
\noindent  For the aim of the exploratory analysis, the information in a tweet can be divided into three main blocks: (i)  general information about the tweet such as the ID, the language, the number of retweets and favorites \footnote{It must be mentioned that the tweets were captured as soon as they were posted, so the values of retweets as favorites are 0.} and especially the body of the tweet (ii) geolocation where the tweet was sent from and  (iii) user information such as  name, description, followers, friends, number of favorite tweets, number of retweets, account location, language, if it is a verified account, and graphic representation of the account.

\section{Methodology }
\label{sec:methodology}

This  paper proposes the application of a reasoning methodology based on prior annotated examples, similar to CBR (Case-Based Reasoning  \cite{1583589}). Classifiers based on CBR determine whether or not an object is a member of a class according to the examples in the base case. The extraction of a formal domain model of the cashtag usage on Twitter is not required, so these classifiers requires less effort in knowledge acquisition when they are compared with rule-based systems, for instance. On building a CBR classifier, first, feature selection and extraction are applied at the base case to remove non-informative features while preserving informative ones (features are extracted from Twitter user activity and posts content on the base case). Second, those features are applied to a classifier that has been trained offline using machine learning techniques.  The specific methodology is summarized in Figure 5 where, after obtaining the dataset and pre-processing it, manual annotation and feature identification is carried out. The former, manual annotation, to obtain heuristic-based classifiers which distinguish cryptocurrency and financial tweets in regulated markets; and the latter, feature identification, to decide about the informativeness of observational variables in the application of machine learning classifiers. Although we are combining traditional feature management and machine learning, the main contribution of our work is applying them for disambiguation, that is, constructing a classifier to disambiguate in terms of features. Also the application field, in the context of the irruption of cryptocurrencies, can be considered a novel contribution.

CNHDS was used to extract the common features of cryptocurrencies tweets while  FTNHDS and AMNHDS were used to extract the common features of company tweets, so that, they are not influenced by the issue of homonym tickers.  In particular, the considered features were (i) pots content (preprocessed  regarding punctuation symbols,  stop words, emoticons and URLs were not considered, also lowercasing and  stemming were applied): most common terms,  most common hashtags, or  number of tickers ; (ii)  user fields: number of followers,  number of favorites, the date for account creation, the date for edition of the default interface, a small description of the account, etc.; and (iii)  place and time of the post (weekday, day time or geolocation). \\

We studied the performance of all the proposed classifiers and also their combined performance by designing \textit{Combined Systems} which use the results from heuristic filtering as an independent variable for the supervised classifiers.  Finally, in a more complex approach, we also studied classification performance by using recurrent neural networks (LSTM - Long Short Tern Memory) networks, which trains an embedding matrix for the most common terms of the dataset, in this case, the 10,000 most common ones, and collects the relative importance of each term and the relationship  between terms. \\

\begin{figure}[ht]
    \begin{minipage}{\columnwidth}
  	\begin{center}

  \includegraphics[width=\columnwidth]{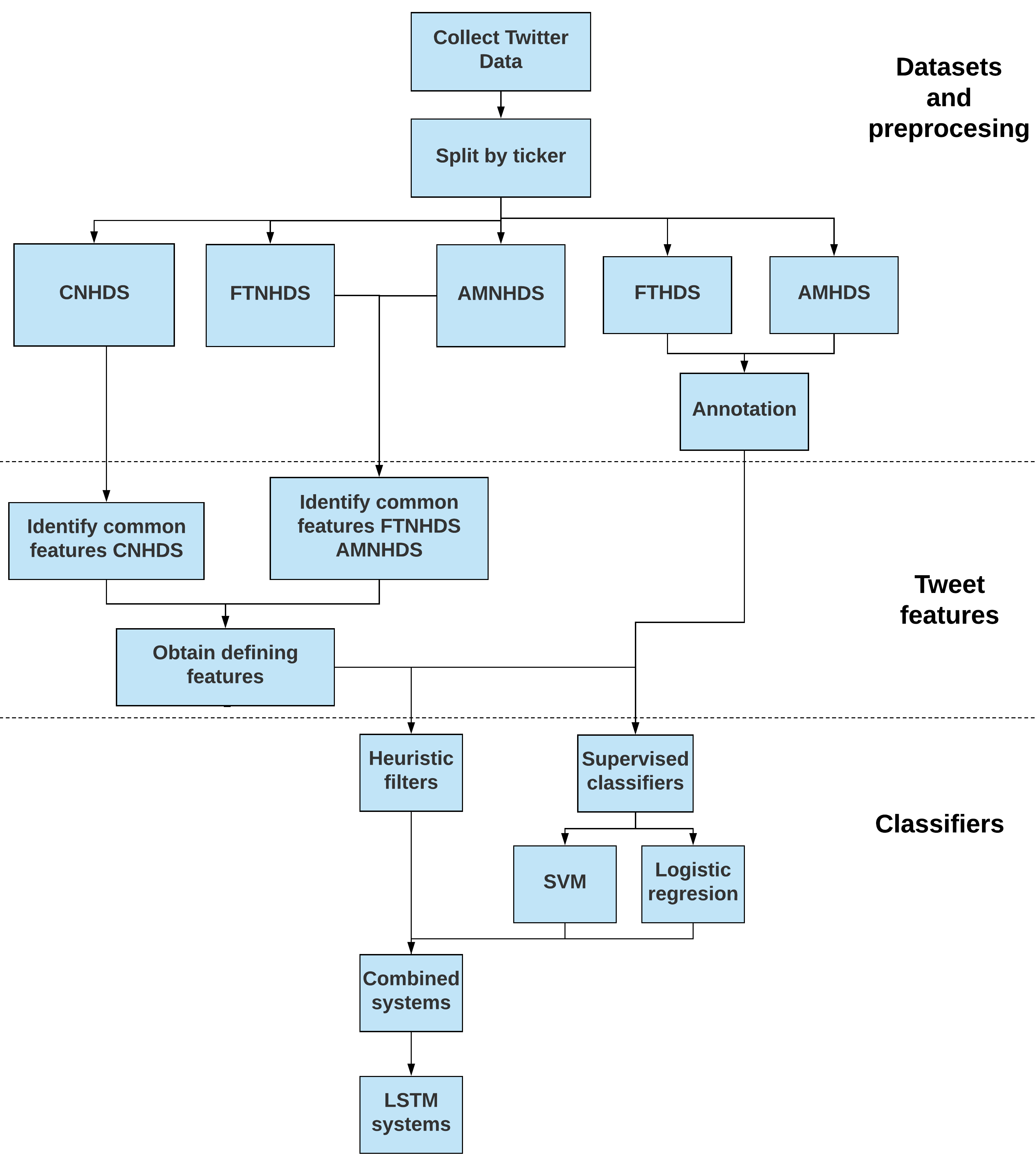}
  \caption{\label{Block diagram} Block Diagram for the proposed method}
  	\end{center}
  \end{minipage}
\end{figure}

\noindent To properly evaluate the proposed classification techniques, the homonym dataset has been randomly split into three subsets: train set (70\%) to  deploy classifiers;  test set (30\%) to measure  their  performance, and tune set (10\% of the train set) to adjust the parameters of the classifiers. We used different performance measures:  precision, recall, specificity, accuracy, F-score and AUC (Area Under the Curve). For instance,  recall -- without totally neglecting precision -- is the key measure for heuristic filtering. However, the F-score and the AUC were the focus for supervised classifiers and combined systems.  In addition to performance, complexity, estimated useful lifespan, updating tasks and usage scope were also evaluated.

\section{Tweet features}
\label{sec:tweet_features}

To identify the defining features from the general information in tweets, FTNHDS and AMNHDS were used as reference to company tweets, and CNHDS as reference for cryptocurrency tweets. Secondly, a user dataset was also built up to take into account the characteristics (number of followers, number of favorites,  date of creation of the account or the edition of the default interface, among others.) of the user who posts, both for cryptocurrency and company tweets. Also, as each user can have a small description on the account, common terms in these descriptions have been also pre-processed (similarly to tweet content) and considered part of the user profile. Thirdly, place (when available)  and time was also considered. In the following sections, although all the tweet fields were observed, only those features which are distinctive according to the type of tweet (company or cryptocurrency) are described. 

\subsection{Content-based features}
\begin{figure}[ht]
  \begin{minipage}{\columnwidth}
	\begin{center}
\includegraphics[width=0.45\columnwidth]{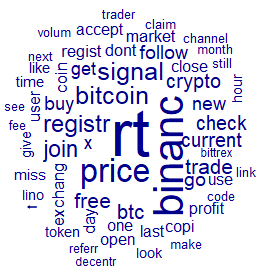} \includegraphics[width=0.45\columnwidth]{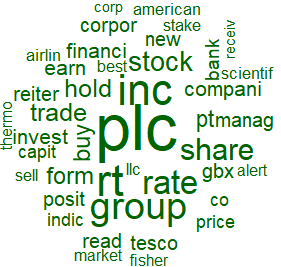}\\
  \caption{Cryptocurrency (blue) \& Company (green) text word cloud \label{cloud}}
	\end{center}
\end{minipage} 
\end{figure}

\begin{figure}[ht]
  \begin{minipage}{\columnwidth}
   \begin{center}
  \includegraphics[width=0.45\columnwidth]{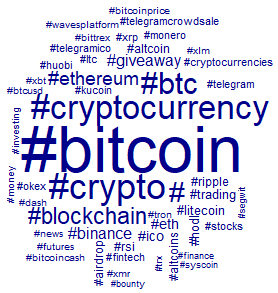}  \includegraphics[width=0.45\columnwidth]{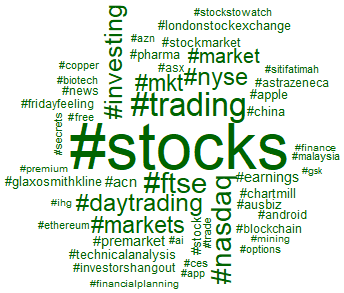}\\
  \caption{Cryptocurrency (blue) \& Company (green) hashtag word cloud \label{hashtag-cloud}}
	\end{center}
	\end{minipage}
\end{figure}

\noindent {\it Most common terms} is the most distinctive feature we can extract from the tweet content since these terms vary from cryptocurrency tweets to company tweets, so that the presence of specific terms can determine the category membership with high likelihood. Figure \ref{cloud} visualizes the feature {\it most common terms} as a tag cloud. The figures show: (1) terms like {\it coin}, {\it crypto}, {\it cryptocurrency}, {\it binanc}, {\it signal}, {\it fee} or {\it join} are defining terms for cryptocurrency tweets while {\it rate}, {\it group}, {\it inc}, {\it plc}, {\it finance} or {\it company} for companies; (2)  the usage of companies and cryptocurrencies proper names is a defining criterion in both cases; and (3)  there area also non-defining terms, mostly referring to market interactions, i.e. {\it buy} so they are considered poor signs of category membership.  Also {\it most common hashtags} can be considered a distinctive feature in the exploratory analysis according to the expert annotation and, in fact, they are similar to  {\it most common terms} (see  Figure \ref{hashtag-cloud}). For instance {\it \#bitcoin}, {\it \#ethereum}, {\it \#cryptocurrency}, {\it \#altcoin}, {\it \#airdrop} or {\it \#binance} for cryptocurrency tweets and {\it \#ftse}, {\it \#mkt}, {\it \#premarket} or {\it \#earnings} for company e tweets. As in {\it most common terms} there are also common hashtags in the two datasets, i.e. \#hold or \#buy  although with very different percentages.\\

\begin{figure}[ht]
  \scalebox{1}[1]{\includegraphics[width=\columnwidth]{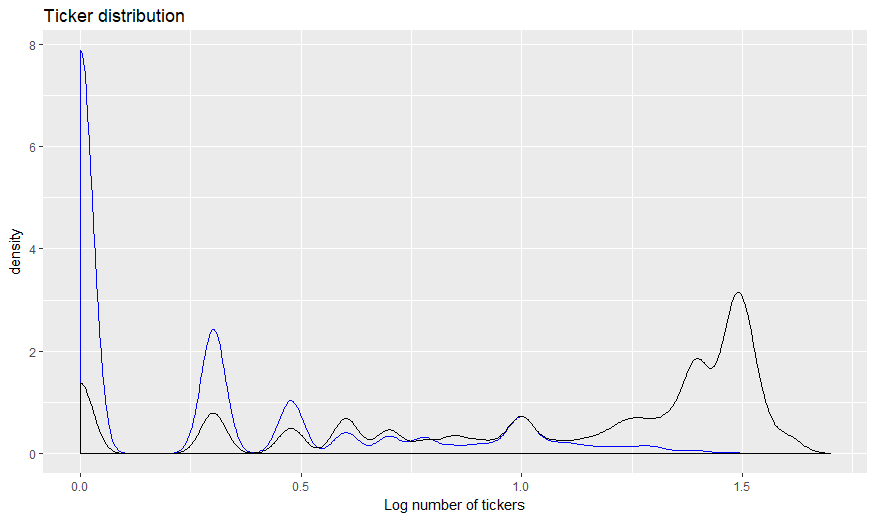}}
  \caption{Ticker distribution, \textit{LSE-100}(blue) vs \textit{Cryptocurrency}(black)\label{TickerDistribution}}
\end{figure}

\noindent Finally, the exploration of the amount of tickers is especially relevant. While 3 is the average of tickers and 1 the median for company tweets, the average and median rise to 18 and 20 respectively for cryptocurrency tweets (Figure \ref{TickerDistribution}. As outliers with a large number of company tickers exist, this feature should not be used in isolation.  \\

\subsection{User information}

\begin{figure}[ht]
  \begin{minipage}{\columnwidth}
	\begin{center}
   \includegraphics[width=0.45\columnwidth]{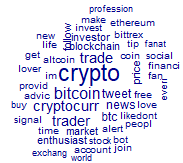}  \includegraphics[width=0.45\columnwidth]{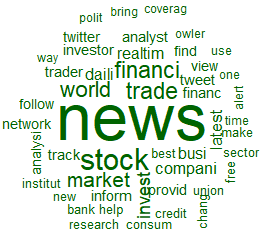}\\
  \caption{Cryptocurrency (blue) \& Company (green)  user description  word cloud \label{UserDescription}}
	\end{center}
	\end{minipage} 
\end{figure}

\noindent  {\it most common terms}  in the description of the user  (Figure \ref{UserDescription}) are quite similar to those extracted from the main body, so i.e. {\it crypto} and  {\it bitcoin} are common for cryptocurrency tweets meanwhile i.e. {\it finance} and  {\it company} are so. Moreover, the user description tends to have less formal and more personal words (i.e. {\it enthusiast} or {\it love}. However, a relevant number  of  common terms are shared between users' description in  cryptocurrency and company tweets i.e. {\it news} or {\it invest}, so the users' descriptions are not as defining as the tweet content in term of category membership.

\begin{figure}[ht]
  \begin{minipage}{\columnwidth}
   \includegraphics[width=\columnwidth]{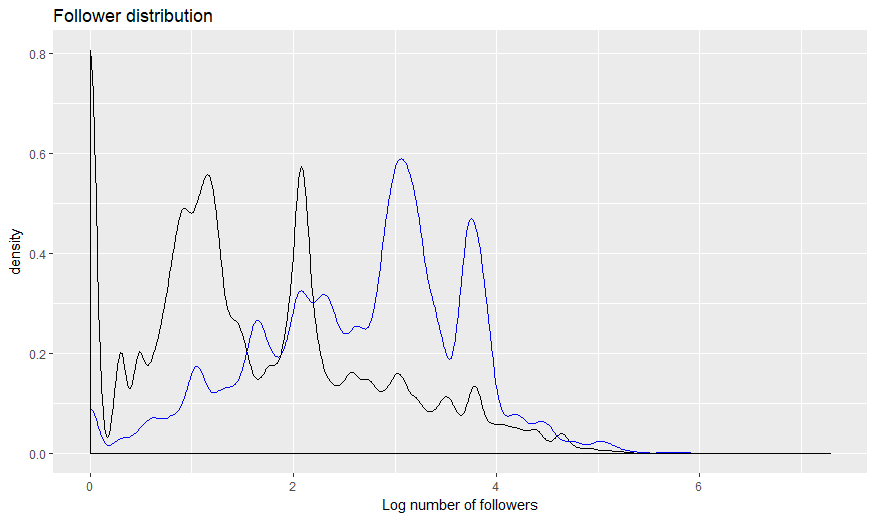}\\
  \caption{Follower distribution by user, \textit{LSE-100} (blue) vs \textit{Cryptocurrency} (black) \label{Followers}}
  \includegraphics[width=\columnwidth]{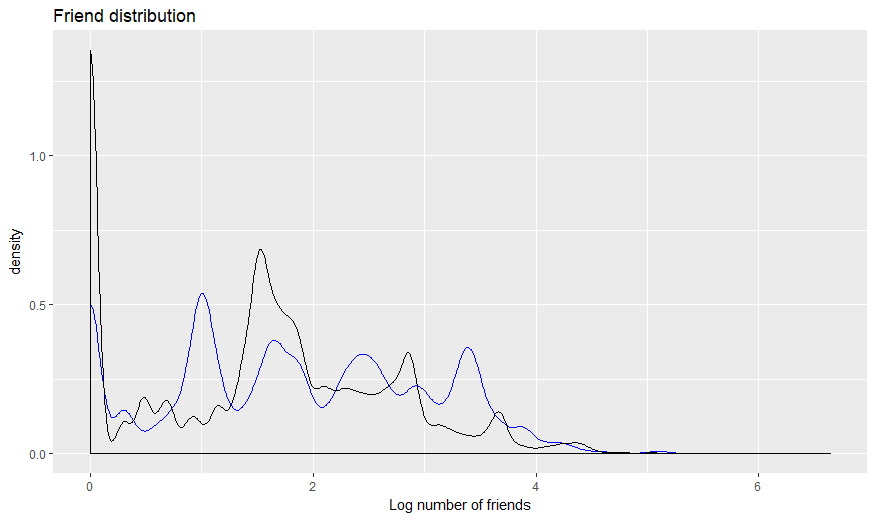}\\
  \caption{Friend distribution by user, \textit{LSE-100} (blue) vs \textit{Cryptocurrency} (black) \label{Friends}}
	\end{minipage} 
\end{figure}

\noindent On the other hand, especially for LSE-100, the number of followers and friends was uncovered as a defining feature (Figures \ref{Followers} and \ref{Friends});  so that followers of users linked to cryptocurrency tweets tend to be quite small (more than three quarters below  200 and most of them below 2 \footnote{We interpret these numbers as a result of the existence of  secondary accounts to disseminate self-generated tweets}) meanwhile the percentage increases for company tweets where above 75\% of the users have at least 100  followers. This defining feature results also on differences in the number of retweets and favorites but they are not as remarkable as the number of followers and friends.  However, even for users linked to cryptocurrency tweets, there are outlier users with millions of subscribers, which shows that the use of cashtags to refer to cryptocurrencies is a widespread phenomenon. 

\noindent Finally, and focusing on the account, its verified status, its change of profile and its creation time were considered relevant during the exploration.   For the verified status, although the percentage of verified users is not very high, 1\% for tweets about companies, they are slightly more frequent than in cryptocurrency accounts (0.1\%), which can be also a result of the greater number of followers found in users linked to company tweets. Secondly,  most accounts  (72\%) linked to cryptocurrency tweets have not changed the default profile (which  is consistent with accounts for non-personal use but for the diffusion of self-generated messages); meanwhile, for instance, in the case of LSE-100 companies, this percentage falls to 58\%. Thirdly, for the account creation time, the accounts linked to LSE-100-company tweets were created between 2009 - 2017; meanwhile,  cryptocurrency accounts are recent, from mid-2017 to the present, a period that coincides with the expansion of cryptocurrencies. However, the defining nature of creation time is reduced in case of recent accounts, so it should be combined with other defining features. \\

\subsection{Tweet time and place}
\begin{figure}[ht]
  \centering 
  \scalebox{1}[1]{\includegraphics[width=\columnwidth]{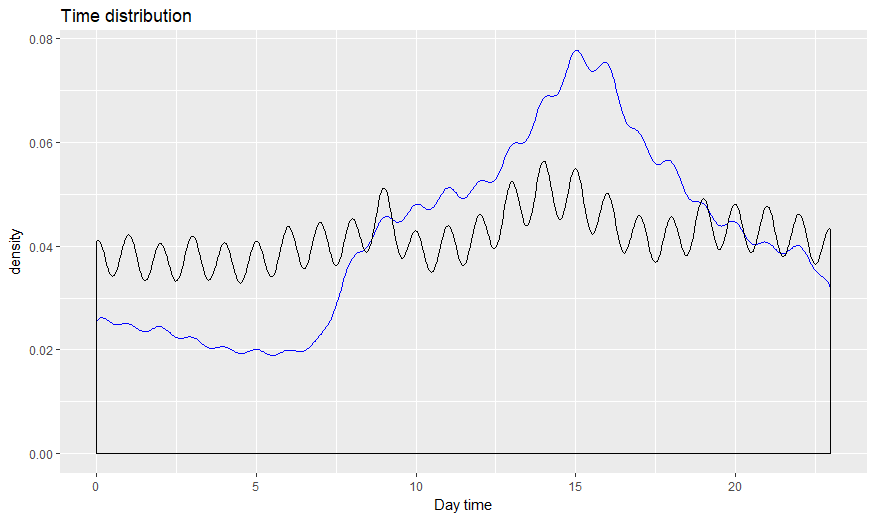}}
  \caption{Tweet time distribution, \textit{LSE-100}(blue) vs \textit{Cryptocurrency}(black) \label{Tweet timedistribution}}
\end{figure}

\noindent During the exploratory analysis, the time of the day with the highest number of tweets is considered the most distinguishing feature (Figure \ref{Tweet timedistribution}). Meanwhile LSE-100 tweets are regularly posted within 10 am and 18 pm GMT, when the stock market is open, cryptocurrency tweets are more stable throughout the day, as they do not have a closing time or a specific geographical area. Regarding the posting location, the percentage of geolocated tweets in the datasets are no significative enough to define a  heuristic. \\

\subsection{Main Features}
\noindent After the exploratory analysis over cryptocurrency and company tweets which contain homonym cashtags, each one has distinctive features to tackle automatic classification. These features are  summarized in Table \ref{Tweet main features}

\begin{table}[ht]
	{\footnotesize
\centering
\begin{tabular} {lll}
  \hline
\multicolumn{3}{c}{\textbf{Tweet features}} \\   
  \hline
\textbf{Feature} & \textbf{Cryptocurrency tweets} & \textbf{LSE-100 tweets} \\
 \hline

\begin{minipage}{0.8 cm} Body\end{minipage}	& 
	\begin{minipage}{3 cm}

\vspace{1 mm}
 Terms like {\it crypto, coin, binanc} or name of cryptocurrencies\\
  Many different tickers in the body\\

\end{minipage}	& \begin{minipage}{3 cm}

\vspace{1 mm}
Terms like {\it group, inc, plc, financ}, or name of markets\\
 Few tickers per body (one or two)\\
\vspace{1 mm}
\end{minipage} \\

 \hline

\begin{minipage}{0.8 cm} User	\end{minipage}	
	& \begin{minipage}{3 cm}
\vspace{1 mm}

 Few followers and friends\\
 Accounts created recently \\
 Informal Description \\
 Verified users (0.1\%)\\

\end{minipage}	& \begin{minipage}{3 cm}

\vspace{1 mm}
 Moderate followers and friends\\
 Accounts created from 2010 to now\\
 Formal Description \\
 Verified users (1\%)\\
\vspace{1 mm}
\end{minipage} \\
 \hline
\begin{minipage}{0.8 cm}  Time place	\end{minipage}
	& \begin{minipage}{3 cm}
\vspace{1 mm}

 Posting during the whole day\\
 No geographic information \\

\end{minipage}	& \begin{minipage}{3 cm}

\vspace{1 mm}
 Posting when the LSE is open\\
 No geographic information\\

\end{minipage} \\
 \hline
\end{tabular}
}

\caption{Tweet main features \label{Tweet main features}}
\end{table}

\section{Heuristic Filters }
\label{sec:filters}

A{\it Simple word-based heuristic filter} based on the presence of certain key terms was deployed to reduce as much as possible the amount of misclassified company tweets by using  terms that identify  almost unmistakably cryptocurrency tweets, such as: {\it cryptocurrency, lumen, ethereum, bitcoin, blockchain or stellar} and also the  cryptocurrencies whose acronyms do not collide with company tickers (Table \ref{cryptocurrency tickers}).

\begin{table}[ht]
\centering

\begin{tabular} {|c|}
  \hline
  \begin{minipage}{0.9\columnwidth}
{\footnotesize
  \vspace{2 mm}
  \$SNT, \$ADA, \$MTH, \$ADX, \$LSK, \$DSR, \$ARK, \$CLOAK, \$TKN, \$DLC, \$DCR, \$KMD,\$IQT, \$ZCL, \$DCY, \$ALIS, \$RBY, \$SYS, \$EXP, \$BCY, \$VEN, \$BCN, \$BLITZ, \$UGT, \$GVT, \$MONA, \$QASH, \$DASH, \$AUR, \$UNO, \$BURST, \$REQ, \$PART, \$TRIG, \$GCR,\$LMC, \$XEM, \$BNB, \$SNGLS, \$BITSILVER, \$PDC, \$ELIX, \$XVG, \$DOPE, \$LEND, \$SNRG, \$NLG, \$ARDR, \$QSP, \$SALT, \$SYNX, \$GRC, \$XDN, \$PIVX, \$DCT, \$WAVES, \$PTOY, \$SIB, \$LTC, \$CPC, \$NAS, \$XMR, \$LOCI, \$ION, \$VSX, \$NXS, \$XMY, \$GBYTE, \$XMG, \$IGNIS, \$ETP, \$BWK, \$FCT, \$DRGN, \$MUE, \$XPM, \$STEEM, \$FTC, \$SPHR, \$DGB, \$DGD, \$SUB, \$VOX, \$MAID, \$RPX, \$AEON, \$XAUR, \$MIOTA, \$CRC, \$BET, \$ENG, \$XVJ, \$POWR, \$STORJ, \$GUP, \$UBQ, \$SBD, \$INFX, \$LGD, \$DYN, \$INFR, \$ONION, \$MANA, \$SLR, \$FUN, \$CURE, \$BITB, \$EMC2, \$XZC, \$IOTA, \$COVAL, \$AGRS, \$PASC, \$DOGE, \$XRB, \$SWT, \$FLDC, \$ZEC, \$NBT, \$XRP, \$ETH, \$RADS, \$ETC, \$PANGEA, \$CLAM, \$PHR, \$APX, \$BTC, \$NEM, \$NEO , \$MYST, \$START, \$ENJ, \$WTC, \$PPT, \$STR, \$ARDOR, \$ITZ, \$BCPT, \$ITC, \$TAAS, \$STRAT, \$SEQ, \$EDG}
  \vspace{2 mm}
  \end{minipage}\\
  \hline
\end{tabular}
\caption{Cryptocurrencies tickers used in {\it Simple Word-based Heuristic filter}\label{cryptocurrency tickers}}
\end{table}

\begin{table}[ht]
{\footnotesize 
\begin{center}
\begin{tabular} {lll}
  \hline
\multicolumn{3}{c}{\textbf{Heuristic filters}} \\   
  \hline
 & \textbf{Basic} & \textbf{Extended} \\
  \hline
Precision (Company)	& 0.551	& 0.609 \\
Recall	& 0.932 & 0.999 \\
Specificity	& 0.980 & 0.983 \\
Accuracy	& 0.978 & 0.983 \\
F-Score	 & 0.692 & 0.757  \\
\end{tabular}
\end{center}
}
\caption{Word-based heuristics: Quality \label{Wordbase heuristic filter measurements}}
\end{table}

\begin{table}[ht]
\centering
{\footnotesize
\begin{tabular} {lll}
  \hline
\begin{minipage} {3 cm} \begin{center} \vspace{1 mm} \textbf{Reference ticker} \vspace{1 mm} \end{center} \end{minipage} & \begin{minipage} {4 cm} \begin{center} \textbf{Word list} \end{center} \end{minipage}\\ 
  \hline
  \hline
\begin{minipage} {3cm} \vspace{1 mm}  General Cryptocurrencies  \end{minipage} & \begin{minipage} {4 cm} \vspace{1 mm}  coin, crypt, btc, lumen, ethereum, bitcoin, whale, stellar, binanc, blockchain  \vspace{1 mm} \end{minipage}\\ 
\hline
\begin{minipage} {3 cm} \vspace{1 mm}  \$NXT(LSE)  \end{minipage} & \begin{minipage} {4 cm} \vspace{1 mm}  plc \vspace{1 mm} \end{minipage}\\ 
\hline
\begin{minipage} {3cm} \vspace{1 mm}  \$NXT(Crypto) \end{minipage} & \begin{minipage} {4 cm} \vspace{1 mm} ignis, ardor, jelurida \vspace{1 mm}  \end{minipage}\\ 
\hline
\begin{minipage} {3cm} \vspace{1 mm}  \$XLM(LSE) \end{minipage} & \begin{minipage} {4 cm} \vspace{1 mm}  xlmedia \vspace{1 mm}  \end{minipage}\\ 
\hline
\begin{minipage} {3cm} \vspace{1 mm}  \$XLM(Crypto)  \end{minipage} & \begin{minipage} {4 cm} \vspace{1 mm}  rocket, moon, \$str, worth, now, trx \vspace{1 mm}  \end{minipage}\\ 
\hline
\begin{minipage} {3cm} \vspace{1 mm}  \$CRW(LSE)  \end{minipage} & \begin{minipage} {4 cm} \vspace{1 mm}  craneware \vspace{1 mm}  \end{minipage}\\ 
\hline
\begin{minipage} {3 cm} \vspace{1 mm}  \$APH(LSE)  \end{minipage} & \begin{minipage} {4 cm} \vspace{1 mm}  weed, fire, emc, cannabis, medical, amphenol, aphria, \$app, \$acb \vspace{1 mm} \end{minipage}\\ 
\hline
\begin{minipage} {3 cm} \vspace{1 mm}  \$BRK(LSE) \end{minipage} & \begin{minipage} {4 cm} \vspace{1 mm}  amz, aapl, twtr, berkshire, buffet, warren, brookline, brooks, oil \vspace{1 mm}  \end{minipage}\\ 
\hline
\begin{minipage} {3 cm} \vspace{1 mm}  \$SKY(LSE)  \end{minipage} & \begin{minipage} {4 cm} \vspace{1 mm} skyline, fox \vspace{1 mm} \end{minipage}\\ 
\hline
\begin{minipage} {3 cm} \vspace{1 mm}  \$GBG(LSE) \end{minipage} & \begin{minipage} {4 cm} \vspace{1 mm}  plc, group \vspace{1 mm}  \end{minipage}\\ 
\hline
\begin{minipage} {3 cm} \vspace{1 mm}  \$AMS(LSE)  \end{minipage} & \begin{minipage} {4 cm} \vspace{1 mm}  hospital, medical \vspace{1 mm}  \end{minipage}\\ 
\hline
\end{tabular}}
\caption{Words used in {\it Extended Word-based Heuristic Filter}\label{words Etended heuristic filter}}
\end{table}

The quality results of the {\it Simple word-based heuristic classifier} are shown in  the first column in Table \ref{Wordbase heuristic filter measurements}, where the precision, although not very high, is much higher than a null model (2.7\%), so the filter is a good option to discard a lot of tweets about cryptocurrencies only loosing a limited fraction of tweets about companies. The terms used in {\it Simple word-based heuristic classifier} are all specific names of the main current cryptocurrencies or words that refer to them, as would be the case of {\it blockchain} or {\it binance}. Therefore, the performance of the filter should be maintained in the medium term and decline gradually as the trendy cryptocurrencies change. To avoid this, the list of cryptocurrencies should be updated periodically. As it uses a fixed list of cryptocurrencies, the filter should obtain similar results working with tickers different than those studied. Although the precision and recall values obtained are significantly better than those of the null model, more than a thousand tweets from companies are misclassified, which differs from the initial objective of the filter to achieve a practically perfect recall. 

Although all the considered terms in {\it Simple word-based heuristic classifier}  refer directly to cryptocurrencies, in some company tweets the cryptocurrencies are named even when the captured ticker does not refer to a cryptocurrency, as would happen for \$BRK in which various tweets would refer to {\it Berkshire Hathaway} while they talk about cryptocurrencies. This is the reason for most of the failures of the heuristic. To avoid this and improve the performance of this filter, it has been optimized, adding a series of different terms depending on the ticker considered (see Table \ref{words Etended heuristic filter}) in a filter referred to as {\it Extended word-based heuristic filter}. This way, if for example the tweet to consider contains the ticker \$NXT, and terms like {\it Ignis} or {\it Ardor} (elements related to the crypto platforms) the tweet will be classified as belonging to cryptocurrencies. However, if the ticker is \$BRK and contains words like Berkshire or Brookline, the tweet will be marked as a company tweet. These specific criterions will have priority over the general ones. So, if they do not coincide, the labelling of the extended filters will be considered.

The classification performance of this filtering system is shown in Table \ref{Wordbase heuristic filter measurements}. The results of the {\it Extended filter} are significantly higher than those of the {\it Simple filter}. The recall of the system has increased to 99.9\% and only seventeen company tweets are misclassified, a value in line with what was sought for this type of filters. The accuracy of the system has also increased slightly thanks to specific knowledge for each ticker. However, since this new filter  takes specific information about a company, it is limited only to the tickers analyzed, and cannot be used for other cases where the interference between company and cryptocurrency happens.

\section{SVM classifiers }
\label{SVM Classifiers}

Although the heuristic filters successfully detect a large number of tweets about cryptocurrencies, adapting some of the patterns seen during the descriptive analysis to these techniques is complex. Thus, supervised methods have been deployed to effectively split both types of tweets,  more specifically, SVMs (Support Vector Machines).  Unlike heuristic filters, SVM-based solutions  try to achieve a tradeoff between precision and recall, i.e  significant improvements in  precision  at the expense of incorrectly classifying some company tweets. Therefore, the fundamental measurement that will be used to evaluate these classifiers will be the F-score, which allows us to compare the performance of the different classifiers deployed. FTNHDS and AMNHDS have been manually annotated to design the SVMs (The three previously  mentioned subsets were used: train set, test set and  tune set).\\

\noindent The first result that should be highlighted is the SVM classifier that uses the differentiating features observed during the comparison of both types of tweets as independent variables.  Within this set of variables, the posting date has been discarded to not limit the filter to the study period. See the variables in the first SVM approach ({\it Simple SVM Classifier}) in Table \ref{Independent variables Simple SVM}.\\

\begin{table}[ht]
\centering
	{\footnotesize
\begin{tabular} {lll}
	
\textbf{Variable}   &  \textbf{Type}  & \begin{minipage} {3.8 cm} \vspace{1 mm} \textbf{Description}  \vspace{1 mm}  \end{minipage}    \\ 
\hline
\hline
  Ticker   &  Factor & \begin{minipage} {3.8 cm}  \vspace{1 mm} Tickers of the different companies \vspace{1 mm}  \end{minipage}\\ 
\hline
  Weekday &  Integer   & \begin{minipage} {3.8 cm}  \vspace{1 mm} Day of the week when the tweet was posted (from 0 to 4) \vspace{1 mm}  \end{minipage}\\ 
\hline
  Hour   &  Integer   & \begin{minipage} {3.8 cm} \vspace{1 mm} Hour of the day when the tweet was posted  \vspace{1 mm}  \end{minipage}\\
\hline
 Followers  &   Numeric  & \begin{minipage} {3.8 cm}   \vspace{1 mm}  Log10 account followers \vspace{1 mm} \end{minipage}\\
\hline
 Friends  & Numeric  & \begin{minipage} {3.8 cm}  \vspace{1 mm}  Log10 account friends \vspace{1 mm} \end{minipage}\\
\hline
 Favorites & Numeric& \begin{minipage} {3.8 cm}  \vspace{1 mm}  Log2 account favorites \vspace{1 mm} \end{minipage}\\
\hline
 Dollars  &  Numeric  & \begin{minipage} {3.8 cm}  \vspace{1 mm} Log2 number of different tickers in the tweet  \vspace{1 mm} \end{minipage}\\
\hline
DefaultProfile  &  Logical & \begin{minipage} {3.8 cm}  \vspace{1 mm} True if the account has not change the default interface \vspace{1 mm} \end{minipage}\\
\hline
 AccountCreationTime  &  Factor & \begin{minipage} {3.8 cm} \vspace{1 mm} Moment when the account was created (divided in half years) \vspace{1 mm} \end{minipage}\\ 
\hline
\end{tabular}}
\caption{Independent variables for {\it Simple SVM classifier} \label{Independent variables Simple SVM}}
\end{table}

\noindent According to the performance measurements in Table \ref{Measurements SVM classifiers}, the precision values obtained are significantly higher than heuristic classifiers, reaching values close to 90\% with a very low reduction of the recall. In addition, the parameters are considered stable regarding temporal variations, which extends the lifespan of the classifier. The only parameter with a clear temporal component is {\it AccountCreationTime} whose application is justified to differentiate accounts created before and after the irruption of cryptocurrencies. Therefore, the performance of the classifier should remain stable in the medium term. To apply {\it Simple SVM}  to other cryptocurrencies, it should be considered that the ticker used to retrieve the tweet is one of the  model parameters so a new model with new tickers' parameters should be developed when a new ticker appears. On the contrary, a slight degradation in the performance could happen.

To improve lifespan, a model applicable to situations different from those considered during this experiment has been deployed. Moreover, and although the {\it Simple SVM} performance is satisfactory, the information about the content of the tweet is not considered with its full potential. In an alternative SVM-based approach, referred to as {\it Extended SVM Classifier}, relevant terms from the exploratory analysis and key terms identified during manual annotation were considered to differentiate cryptocurrency and company tweets. A specific vocabulary (Table \ref{vovabulary extended classifier}) has been created from these terms and this vocabulary was used to enrich the information through new independent variables ($1$ if the word is in the tweet, no matter how many times, and $0$ otherwise).  A slight improvement in all measurements can be appreciated in Table \ref{Measurements SVM classifiers}, especially accuracy and F-score and even more remarkable in terms of AUC  with a value practically equal to 1 even in the test set. Finally, the {\it Extended SVM} classifier provides precision values greater than 95\% while maintaining a recall higher than 90\%. In terms of lifespan and applicability to scenarios different from the one in this experiment, the consideration of terms in the body of the tweet improves the useful life of the classifier without retraining it, since these words refer to cryptocurrencies and companies and not to specific temporal situations, i.e.  results are considered stable in the medium term. Likewise, the terms in the vocabulary are mainly general and do not refer to specific cryptocurrencies. Even though, few terms in the vocabulary are related to specific companies or cryptocurrencies, e.g. {\it ardor}. There is some little chance of declining performance if the {\it Extended SVM} classifier is applied in another time period. Finally, it is worth to mention that both the execution (classification) and especially the training is slower than {\it Simple SVM} classifier due to the consideration of the vocabulary as an independent variable which results in a bigger number of support vectors.\\

\begin{table}[ht]
{\footnotesize
\begin{tabular} {|c|}
  \hline
  \begin{minipage}{0.9\columnwidth}
  \vspace{2 mm}
Binac, Bitcoin, Signal, Join, Crypto, Fee, Plc, Inc, Group, Company, Finance, Weed, Aapl, Moon, Cannabis, berkshire, Brooks, Ltc, Eth, Dash, Xrp, Xmr, Xem, Nem, Rocket, Jelurida, Ignis, Medical, Buffet, Warren, Stellar
\vspace{2 mm}
  \end{minipage}\\
  \hline
\end{tabular}}
\caption{Vocabulary for the Extended SVM classifier \label{vovabulary extended classifier}}
\end{table}

\begin{table}[ht]
\centering
\begin{tabular} {lll}
	  \hline
	  \multicolumn{3}{c}{\textbf{SVM classifiers}} \\   
	    \hline
 & \textbf{Basic} & \textbf{Extended} \\
  \hline
Precision (Company)	& 0.898	& 0.941 \\
Recall	& 0.897 & 0.935 \\
Specificity	& 0.997 & 0.998 \\
Accuracy	& 0.995 & 0.997 \\
F-Score	 & 0.898 & 0.938  \\
AUC	 & 0.977 & 0.997 \\
\end{tabular}
\caption{SVM classifiers: Quality \label{Measurements SVM classifiers}}
\end{table}

\section{Combined Classifiers }
\label{Combined Classifiers}

In view of the results seen so far, heuristic filters and  SVM classifiers can be used together to supplement each other  benefits. In this section, {\it Combined Systems} are introduced where the results of {\it Extended word-based heuristic filter} is considered  a parameter for the {\it Extended SVM classifier}. The high recall of the former allows a large number of cryptocurrency results to be discarded quickly so that the SVM can focus on precision and, therefore, the combined systems is expected to improve in both metrics. In fact,  precision, recall and F-score values close to 0.97 in the test set (Table \ref{Quality Combined System}) and AUC also improves slightly. Therefore, almost all of the tweets are positively classified, misclassifying only a small percentage. \\

\begin{table}[ht]
	\centering
	{\footnotesize
\begin{tabular} {lll}
  \hline
\multicolumn{3}{c}{\textbf{Combined Classifiers }} \\   
  \hline
  & \textbf{Original} & \textbf{Independent}\\
  \hline
Precision (Company)	& 0.976 & 0.933\\
Recall	& 0.968 & 0.855 \\
Specificity	& 0.999 & 0.998\\
Accuracy	& 0.999 & 0.995  \\
F-Score	 & 0.972 & 0.892 \\
AUC	 & 0.9994 & 0.988\\
\end{tabular}} 
 
\caption{Combined Classifiers: Quality  \label{Quality Combined System}}
\end{table}

\noindent The lifespan and applicability of the {\it Combined Systems} are identical to the SVM classifiers above; the results should remain stable in the medium term, but the benefits will be slightly lower if they are applied to other coincident tickers not included. However, the execution time is slower than SVM classifiers since it requires the consecutive execution of heuristic filters and SVM classifiers. Finally, the working point of the system can be adjusted to obtain slightly higher values of precision or recall depending on the needs.

Although the results are stable in the medium term, the potential to apply the combined system in other scenarios (different from this experiment) is smaller due to the features of some of the variables used in the {\it Extended word-based heuristic} filter. Therefore,  a combined system from the results of {\it Simple word-baed Heuristic} filter (only general information)  can be more stable because, instead of words related to specific cashtags, the basic filter is used. To produce a combined system as stable as possible,  captured ticker information and terms related to specific tickers have been discarded in the {\it Extended SVM classifier}: the variables and vocabulary  for the new {\it Independent Combined System} are shown in \ref{Variables Independent}  and \ref{vocabulary Independent}. The generalization process enables  a classifier easily applicable to scenarios of colliding company tickers and cryptocurrencies  similar to the one studied in this experiment, with a similar performance in the test set (see Table \ref{Quality Combined System}). Although a slight fall can be observed, especially noticeable for recall, precision continues still high, exceeding 90\%; accuracy is greater than 99\%; and AUC  remains high, which allows adjusting other solutions that optimize the precision or recall depending on the desired features. The {\it Independent Combined System} classifier does not use variables with high temporal variability, as the {\it Combined System}, so the results should be stable in the medium term.\\

\begin{table}[ht]
	\begin{minipage}{\columnwidth}
{\footnotesize
\begin{tabular} {lll}
 \textbf{Variable}  &  \textbf{Type}  & \begin{minipage}{3.5cm} \textbf{Description}   \end{minipage}  \\ 
\hline
\hline
 Weekday  & Integer & \begin{minipage}{3.5cm} \vspace{0.1cm} Day of the week of posting (from 0 to 4) \vspace{0.1cm}\end{minipage}\\ 
\hline
 Followers &  Numeric & \begin{minipage}{3.5cm} \vspace{0.1cm} Log10 account followers \vspace{0.1cm}\end{minipage}\\
\hline
 Friends  &  Numeric &  \begin{minipage}{3.5cm} \vspace{0.1cm} Log10 account friends \vspace{0.1cm}\end{minipage}\\
\hline
 Favorites  &  Numeric  & \begin{minipage}{3.5cm} \vspace{0.1cm} Log2 account favorites \vspace{0.1cm}\end{minipage}\\
\hline
Dollars  &  Numeric &  \begin{minipage}{3.5cm} \vspace{0.1cm} Log2 number of different tickers  \vspace{0.1cm}\end{minipage}\\
\hline
DefaultProfile  &  Logical  &  \begin{minipage}{3.5cm} \vspace{0.1cm} True for accounts with the default  interface \vspace{0.1cm}\end{minipage}\\
\hline
 AccountCreationTime  &  Factor  &  \begin{minipage}{3.5cm} \vspace{0.1cm} Account creation time  (divided in half years)\vspace{0.1cm} \end{minipage}\\ 
\hline
\end{tabular}}
\caption{Variables in {\it Independent Combined Classifier} \label{Variables Independent}}
	\end{minipage}
\end{table}

\begin{table}[ht]
\centering
\begin{tabular} {|c|}
  \hline
  \begin{minipage}{0.9\columnwidth}
  \vspace{2 mm}
{\footnotesize 
Binac, Bitcoin, Signal, Join, Crypto, Fee, Plc, Inc, Group, Company, Finance, Aapl, Moon, Ltc, Eth, Dash, Xrp, Xmr, Xem, Nem, Rocket}
\vspace{2 mm}
  \end{minipage}
  \\
  \hline
\end{tabular}
\caption{Vocabulary in {\it Extended SVM Classifier} for the {\it Independent Combined Classifier} \label{vocabulary Independent}}
\end{table}

\section{LSTM classifiers }
\label{LSTM Classifiers}

The heuristic filters, SVM classifiers and the combined classifying systems proposed so far consider a set of terms as a relevant representation of the  tweet content. However, none of them considers the relative importance of each of these terms or the relationship that may exist among them. For this reason,   the aforementioned combined classifying systems have been adapted to work according to LSTM (Long short-term memory network) classifiers, as mentioned, a type of recurrent network. Thus, instead of a set of terms, an embedded matrix that collects the relative importance and inter-relationship of terms is used in the classifier. Moreover, it is fair to mention that LSTM adaptation allows a greater number of terms in the vocabulary without an excessive increase in the number of independent variables.\\

\noindent FTHDS and AMHDS have been used as input to an LSTM network in order to obtain the embedding matrix. In particular, an LSTM network  aims to predict the next word  from a set by considering the previous words. For this, a matrix is generated which includes the weight of the considered term (as a measure of its relevance for the problem) and the relationships among  terms. The matrix together with the network are iteratively trained so a suitable number of terms should be defined to guarantee and affordable computational complexity. In our experiment, to obtain an {\it LSTM-based classifier},  tweets are represented as  vectors and the vocabulary consists of the 10,000 most common terms within the homonymous tweets. Before LSTM training, a pre-processing step is applied, which includes the following tasks: (1)  removing weird characters, punctuation, emoticons, URLs and stop words; (2) discarding tickers and names of cashtags and extremely common terms; and (3) stemming. After this preprocessing step, the resulting terms have a higher representative capacity and the final vocabulary consists of the 9,998 terms, in addition to a term for those terms not collected and another for the break line. After training the LSTM network, the matrix together with the tweets' vectors are used  to generate new independent variables by multiplying tweets' vectors by the LSTM matrix. As a result, in addition to a significant reduction in the number of variables (from 10,000 to 200), a more relevant (in terms of the problem)  representation of the tweet body is obtained. To sum up, for the LSTM approach in our experiment, a tweet is represented by a vector of 200 variables, which are the input independent variables of the SVM classifier. \\

\begin{table}[ht]
\centering
{\footnotesize
\begin{tabular} {lll}
  \hline
 \multicolumn{3}{l}{\textbf{LSTM-SVM Classifiers}} \\   
  \hline
  & {\bf Original} & {\bf Independent}\\
   \hline
Precision (Company)	& 0.981 & 0.967 \\
Recall	& 0.969  & 0.928 \\
Specificity	& 0.9995 & 0.999 \\
Accuracy	& 0.999  & 0.997  \\
F-Score	 & 0.975  & 0.947 \\
AUC	 & 0.9990 & 0.992 \\
\end{tabular}}  

\caption{ LSTM-adapted SVM: Quality  \label{Measurements AUC LSTM SVM classifier}}
\end{table}

\noindent The {\it Combined Classifiers} in Section \ref{Combined Classifiers} are deployed again but using the results of the embedding matrix instead of the list of common terms (basic or extended). As shown in Table \ref{Measurements AUC LSTM SVM classifier}, there are no major changes in performance. Since the extra computational load in LSTM-SVM classifier does not provide a significant improvement in performance (slight improvement from previous classifiers) is not worthy to be considered as an isolated solution, however, LSTM can produce performance improvements in the combined classifiers. On the other hand, limitations and applicability of the {\it LSTM-adapted SVM classifier} are the same as SVM classifiers. The {\it LSTM-adapted SVM classifier} may show a small drop in its performance when used in scenarios with companies different from the one considered in this experiment; but  {\it LSTM-adapted SVM classifier}  maintain the performance in the medium term for the same company scenario since  variables with a fairly clear temporal variation are not in the model.\\

\noindent Unlike the {\it LSTM-adapted SVM} classifier, the {\it Independent LSTM-adapted SVM classifier} provides large improvements compared to the same model with key terms. These improvements are especially noticeable for the precision, recall and F-measure of the system, surpassing 0.92 for all of them, unlike the 0.855 of the previous independent classifier. The significantly greater vocabulary used increases the representative capacity and compensates the reduction in the other fields. In fact, the benefits obtained are similar to those of the {\it LSTM-adapted SVM} classifier, which virtually classifies correctly all tweets. For this reason, the use of this model would be advisable to process homonym tickers different from those studied, although with a computational overload due to the large size of the vocabulary, embedding matrix and support vectors. As with the previous classifiers, it does not use variables with high temporal variability, so the results obtained should be maintained in the medium term. To maintain long term benefits, it would be necessary to update the temporary matrix every few months to adapt to changes in the new terms used. However, given the large size of the vocabulary, most of them should not change. So, the performance of the system should reduce more slowly than the independent classifier.\\

\section{Logistic-regression-based Classifiers }
\label{Logistic-regression-based Classifiers}

Despite the good results obtained through the precious  SVM classifiers, the execution and especially the training of SVMs can be slow and, more importantly,  they are grey-box models so hard to interpret their results to further improvements. As a  white-box alternative, a set of {\it logistic-regression-based} classifiers were considered, which, in addition, are simpler and so faster. If similar quality results can be obtained with a simpler model, the extra complexity may be unjustified. In our experiment, SVM was replaced by logistic regression but, given the high computational cost of the LSTM network, it has not been used in the logistic-regression-based models. Keep in mind that the aim is obtaining a simpler and white-box solution. See the quality results of these classifiers in Table \ref{quality Logistic-regression-based}.

\begin{table}[H]
	\begin{minipage}{\columnwidth}
{\footnotesize
\begin{tabular} {lllll}
  \hline
\multicolumn{5}{l}{\textbf{LR Classifiers}} \\   
  \hline
 & \textbf{Basic} & \textbf{Ext.} & \textbf{Comb.} & \textbf{Ind.} \\
   \hline
Precision (Company)	& 0.816 & 0.914 & 0.950 & 0.871\\
Recall	& 0.807 & 0.872 & 0.960 & 0.801 \\
Specificity	& 0.995 	& 0.998 	& 0.999 	& 0.997\\
Accuracy	& 0.990 & 0.994 & 0.998 & 0.991 \\
F-Score	 & 0.812 & 0.892 & 0.955 & 0.835  \\
AUC	 & 0.977 & 0.993  & 0.9997  & 0.986\\
\end{tabular}
}

\caption{ Logistic-regression-based: Quality \label{quality Logistic-regression-based} {\footnotesize (Basic, Extended, Combined, Independent)}}

	\end{minipage}
\end{table}

\noindent Although the quality of all the logistic regression models are lower than SVM-based classifier, the fall is not very significant, the execution time of these logistic-regression-based classifiers is significantly lower (five times faster and more). The basic logistic regression classifier is especially noteworthy since the tweet content does not need to be processed so it can be trained and applied really fast. Regarding limitations,  {\it logistic-regression-based Classifiers} maintain similar constraints and restrictions as SVM-based classifier  because the same independent variables are used.\\

\section{Limitations }
\label{limitations}

In this experiment, different alternatives have been explored  to disambiguate homonyms terms in the LSE-100 and in the cryptocurrency market with the final aim of clearly distinguish tweets referring to companies in regulated markets (LSE-100  in the experiment) and tweets regarding cryptocurrencies and so in a not regulated market. First, word-based heuristic filters have the main benefit of discarding  a large number of cryptocurrency tweets without practically miss-classifying any company tweet, so they achieve high recall values with acceptable levels of precision. Secondly, classifiers based on supervised methods provide a tradeoff between precision and recall, maximizing the F-score quality measure. For both  heuristic filters and supervised classifiers, different alternatives have been explored and analyzed in terms of quality measures for binary classification and in terms of computational load. High-quality results have been obtained for the more complex and computationally expensive models.\\

\noindent In a second step and from the supplementary benefits of heuristic filters and SVMs, we have explored the combined deployment of both types of classifying models. As a result, we obtain a {\it Combined System} which AUC values very close to $1$ and F-score above 0.975. Thirdly, classifiers able to identify company and cryptocurrency tweets that do not use information related to any of the studied cryptocurrencies are also studied.  These {\it Independent Models}, despite a small decrease in classification quality, still maintain high levels of precision and recall, especially if they use an LSTM embedding matrix instead of a fixed list of key terms. These cryptocurrency-independent models offer the potential to be used in scenarios different from the experiment in this paper. Also, their working points in AUC can be adjusted to the best option in a specific problem. Finally, logistic regression as a less computationally expensive and a more interpretable solution has been applied in the experiment, and especially for the case of {\it Extended logistic regression classifier}, a quick initial classification scheme can be deployed. \\

\noindent Regarding the limitations of the developed classifier systems, two cases should be differentiated. In the first place, there would be those classifiers that use information that refers to some of the tickers considered in the experiment, such as {\it Extended  Word-based  Heuristic Filter}, {\it Extended SVM Combined System},  or  {\it LSTM-adapted SVM Classifier}. They use input variables as company tickers or key terms related to some of the cashtag in the experiment  to achieve an improvement in classification quality.  This means that their performance for company tickers not in the experiment may be lower. Thus,  they are especially suitable to work with the companies of the LSE-100 but their benefits fall outside this stock index.\\

\noindent In the second place, classifying systems, which do not use  information regarding the cashtags in the experiment, maintain their performance in scenarios out of the  tickers in the experiment (independent classifiers) like  {\it LSTM independent classifier} or {\it simple word-based Heuristic Filter}. For these classifiers, the quality performance is stable in the medium term, since information with time-dependent nature is not considered apart from the account creation time. However,  creation time is merely included to differentiate accounts created before and after the popularization of cryptocurrencies. Thus, they should continue to work properly. However,  firstly, popular cryptocurrencies may vary from time to time and the list of cryptocurrency tickers should be updated every few months to maintain the classifying performance of the system; and secondly  common terms form tweet content may also vary. To sum up, the embedding matrix should be re-computed every few months to keep the classifying performance of the system. The other independent variables  considered should maintain regular behavior, at least in the medium term.\\

\section{Model Evaluation and Selection}
\label{modelEvaluation}

Given that Independent Classifiers are considered superior in term of generalization to other markets, in this section, the three independent models are further evaluated to check  whether there are a statistical difference among them: Independent SVM Combined Classifier (SVM-Ind), Independent LTSM-adapted SVM Classifier (LSTM-Ind)  and Independent Logistic-Regression based Classifier (LR-Ind). Non-parametric statistical tests are applied to SVM-Ind, LTSM -Ind and LR-Ind \cite{DBLP:journals/corr/abs-1811-12808}: (1) The McNemar test (for paired comparisons) which compares the performance of two machine learning classifiers; and (2) The Cochran’s Q test as a generalized version of McNemar’s test that can be applied to compare three or more classifiers.

The Cochran's Q test is a nonparametric statistical test to evaluate the null hypothesis. If the test result suggests that there is insufficient evidence to reject the null hypothesis, then any difference observed in the performance of the models is probably due to statistical chance. Conversely, if the test rejects the null hypothesis, it is likely that the different performances are due to a difference in the models.

Let $\{C_1, \dots , C_M\}$ be a set of classifiers who have all been tested on the same dataset. 

If the $M$ classifiers do not perform differently, then the following $Q$ statistic is distributed approximately as {\it chi squared} with $M-1$ degrees of freedom:

$$Q = (M-1) \frac{M \sum^{M}_{i=1}G_{i}^{2} - T^2}{MT - \sum^{N_{ts}}_{j=1} (M_j)^2}$$

Here, $G_i$ is the number of objects out of $N_{ts} \times M$ correctly classified by $C_i= 1, \dots  M; M_j$ is the number of classifiers out of M that correctly classified object $$\mathbf{z}_j \in \mathbf{Z}_{ts}, where \mathbf{Z}_{ts} = \{\mathbf{z}_1, ... \mathbf{z}_{N_{ts}}\}$$ is the test dataset on which the classifiers are tested on; and T is the total number of correct number of votes among the M classifiers:

$$T = \sum_{i=1}^{M} G_i = \sum^{N_{ts}}_{j=1} M_j$$

The Cochran’s Q test can be considered as a generalized version of the McNemar’s test, which is applied to compare the predictions of two models to each other to evaluate the null hypothesis.

$$\chi^2 = \frac{(B - C)^2}{(B + C)}$$

where B and C are the predictions in which the two models differ: one made a correct prediction an the other an incorrect prediction, or vice versa.

In this study, both test are applied to the three Independent Models, as a result, we get the $Q$ and $\chi^2$ values shown in Table~\ref{tab:q_chi2_all_dataset}.

\begin{table}[h] \centering

\begin{tabular}{ccc} \cline{1-3} \textbf{} &\textbf{Q | $\chi^2$ } &\textbf{$p$-value} \\ \hline

\textbf{Cochran’s Q }& 678.135 & 5.56e-148 \\
\textbf{McNemar’s SVM- Ind vs LR-Ind}& 317.071 & 6.29e-71 \\ 
\textbf{McNemar’s SVM-Ind  vs LSTM-Ind}& 109.835 & 1.07e-25\\ 
\textbf{McNemar’s LSTM-Ind  vs LR-Ind}& 489.507 & 1.82e-108\\ \hline

\end{tabular} \caption{Cochran’s Q and McNemar’s test (all dataset)} \label{tab:q_chi2_all_dataset} \end{table}

To avoid the effect of a test set that is too large \cite{p-value-1, p-value-2}, we divide the data into subsets of 10,000 random samples and calculate the average of the values obtained. As a result, we get the Q and $\chi^2$ values shown in Table~\ref{tab:q_chi2}.

\begin{table}[h] \centering

\begin{tabular}{ccc} \cline{1-3} \textbf{} &\textbf{Q | $\chi^2$ } &\textbf{$p$-value} \\ \hline

\textbf{Cochran’s Q }& 35.239 & 5.99e-05 \\
\textbf{McNemar’s SVM-Ind vs LR-Ind}& 15.528 & 0.00333 \\ 
\textbf{McNemar’s SVM-Ind vs LSTM-Ind}& 6.991 & 0.0744 \\ 
\textbf{McNemar’s LSTM-Ind vs LR-Ind}& 25.184 &7.82e-05 \\ \hline

\end{tabular} \caption{Cochran’s Q and McNemar’s test} \label{tab:q_chi2} \end{table}

In view of the result of the Cochran’s Q test, assuming a significance level of $\alpha=0.05$, we can reject the null hypothesis since the corresponding p-value is lower. However, the McNemar’s test gives us further information when comparing the two-to-two models: we can reject the null hypothesis between the SVM-Ind and LSTM-Ind models versus the LR-Ind model, but not between the SVM-Ind model and the LSTM-Ind model. Consequently, we can conclude that the improvements in performance obtained with the Independent LSTM-adapted SVM classifier in our study may be due to statistical chance, so it is not worth the greater complexity and computational cost it requires.

\section{Conclusions and future lines }
\label{conclusions}

Despite  cashtags are the main mechanisms to track financial information on Twitter, the  irruption of the cryptocurrencies  has produced a degradation in the quality of the information obtained through cashtag tracking due to the fact that some cryptocurrency acronyms collide  with company tickers (homonym tickers) in regulated markets. When a cashtag is used on Twitter, homonym acronyms can extract results referring to stock companies and to cryptocurrencies indistinctly. In addition, most cryptocurrency tweets are self-generated spam messages that multiply the negative effect of the homonyms acronyms with a bid degradation of the informative capacity the cashtag seeks to obtain. Thus, new disambiguation mechanisms -or the adaptation of existing ones- are necessary to solve the problem and restore the original informative capacity of the cashtag tracking mechanism.\\

\noindent Meanwhile, most of the current researches are focused on the potential of Twitter as a predictive tool for decision making on financial markets and the development of  expert systems  using such information, the approach of this paper is focused on a completely different objective: to illustrate the negative impact of the increasing  popularity of cryptocurrencies in the cashtag mechanism through an experiment on LSE-100 companies. The aim is the deployment of classifying systems able of differentiating company and cryptocurrency.\\

\noindent Based on these features, different classifying systems have been introduced  to identify or distinguish cryptocurrency and company tweets for the case of homonyms acronyms. {\it Word-based  Heuristic Filters} pursue discarding a large number of cryptocurrency tweets without practically misclassifying company tweets, so they achieve high recall values with acceptable levels of precision. On the other hand, classifiers based on supervised methods provide a tradeoff between precision and recall by maximizing F-Score and high-quality results have been obtained for the more complex and computationally expensive models. In addition, we have analyzed the combined action of heuristics and supervised models, which results in an approach reaching AUC values very close to $1$ and F-score above  $0.97$. \\

\noindent Regarding applicability and the ability to update to other scenarios, classifiers that do not use information related to any of the studied cryptocurrencies have been also studied. These models, despite a slight decrease in classifying metrics, still preserve a high level of precision and recall, especially when they do not use a fix list of key terms but a dynamic LSTM matrix. This good performance opens the possibility to use these solutions in different scenarios from the one studied in this experiment. Finally, the work point of the model in AUC can be adjusted according to the problem needs, a simple solution based on logistic regression classifier can be used as an initial classifier to obtain a quick estimate for the classification.\\

\noindent During this paper, the influence of cryptocurrency tweets in the cashtag results is analyzed for the main LSE-100 companies. Although during the study period, from July 1 2017 to February 15 2018, the interference between the cryptocurrencies and  tickers of LSE-100 only appeared for the cashtags indicated in our experiment, recently the negative impact of the interference has grown, eg. \textit{\$SPH} (Sinclair pharma (AIM-100) vs Sphere(coin)), \textit{\$REDD} (Redde (AIM-100) vs Reddcoin) and \textit{\$SMT} (Scottish mortgage investment trust plc(FTSE-100) vs SmartMesh(coin)).  All these new cryptocurrencies  increased their popularity highly before the study period. On the other hand, homonym tickers between cryptocurrencies and stock companies are also found in other markets such as the NSQE or NASDAQ.  As the Independent Models are the most detached from training data, they have the greatest potential to be used in  different stock markets so that they provide trans-applicability meanwhile they retain performance in other contexts. Nonetheless, our future work also addresses testing the applicability of independent classifiers in up-to-date LSE-100 scenarios and other regulated stock markets to check the performance in the presence of other colliding cashtags.  This applicability testing will also pursue to measure the adaptation cost of non-independent classifiers to these new cases.

\bibliographystyle{elsarticle-harv}

\end{document}